\documentclass[aps,prl,twocolumn,superscriptaddress,longbibliography]{revtex4-1}

\usepackage{mathrsfs}
\usepackage{tabularx}
\usepackage{slashed}
\usepackage{amsmath}
\usepackage{amsfonts}
\usepackage{graphicx,color}
\usepackage{times}
\usepackage[caption=false]{subfig}
\usepackage[colorlinks,linkcolor=red,citecolor=blue]{hyperref}
\newcommand{\comment}[1]{}
\AtBeginDocument{%
    \newwrite\bibnotes
    \def\bibnotesext{Notes.bib}
    \immediate\openout\bibnotes=\jobname\bibnotesext
    \immediate\write\bibnotes{@CONTROL{REVTEX41Control}}
    \immediate\write\bibnotes{@CONTROL{%
    apsrev41Control,author="08",editor="1",pages="1",title="0",year="1"}}
     \if@filesw
     \immediate\write\@auxout{\string\citation{apsrev41Control}}%
    \fi
}%

\begin{document}

\title{Quantum Metric Induced Phases in Moir\'e Materials}

\author{Ahmed Abouelkomsan}
\thanks{A.A and K.Y. contributed equally to this work}
\author{Kang Yang}
\thanks{A.A and K.Y. contributed equally to this work}
\author{Emil J. Bergholtz}
\affiliation{Department of Physics, Stockholm University, AlbaNova University Center, 106 91 Stockholm, Sweden}

\date{\today}
\begin{abstract}
    We show that, quite generally, quantum geometry plays a major role in determining the low-energy physics in strongly correlated lattice models at fractional band fillings. We identify limits in which the Fubini-Study metric dictates the ground states and show that this is highly relevant for Moir\'e materials leading to symmetry breaking and interaction driven Fermi liquids. This phenomenology stems from a remarkable interplay between the quantum geometry and interactions which is absent in continuum Landau levels but generically present in lattice models where these terms tend to destabilize e.g. fractional Chern insulators. We explain this as a consequence of the fundamental asymmetry between electrons and holes for band projected normal ordered interactions, as well as from the perspective of a self-consistent Hartree-Fock calculation. These basic insights about the role of the quantum metric turn, when dominant, an extremely strongly coupled problem into an effectively weakly coupled one, and may also serve as a guiding principle for designing material setups. We argue that this is a key ingredient for understanding  symmetry breaking phenomena recently observed in Moir\'e materials. 
\end{abstract}

\maketitle

The application of geometry in physics continues to stimulate new fundamental insights. One of the most prominent examples is the general relativity. In condensed matter physics, the role of geometry has been in the limelight since the discovery of the geometric Berry phase \cite{berry1984quantal,PhysRevA.36.3479}, which is the phase accumulated during an adiabatic evolution. The Berry phase has proven to be critical to topological states and transport properties \cite{kohmoto1985topological,PhysRevLett.93.206602,PhysRevB.70.205338,RevModPhys.82.1959,goerbig2012,PhysRevB.85.241308,jackson2015geometric,annurev-conmatphys-031016-025458,PhysRevB.101.035411,varjas2021topological,PhysRevB.102.165148,parker2021field}. Quantum states also trace out a distance during an adiabatic evolution, which is captured by a metric \cite{provost1980riemannian,PhysRevLett.65.1697}. The geometric concept of distance has been well recognized in quantum information theory \cite{PhysRevA.54.1844,PhysRevLett.78.2275,PhysRevLett.113.140401}, and it has also begun to attract interest also in condensed matter physics \cite{PhysRevB.90.165139,PhysRevLett.114.236802,PhysRevB.94.134423,rhim2020quantum,PhysRevB.104.115160,PhysRevB.104.L180502,PhysRevB.102.165118,PhysRevB.105.085154,northe2021interplay,PhysRevB.97.201117,PhysRevResearch.1.032019,PhysRevLett.122.210401,10.1093nsrnwz193}. 
Examples include the collective excitations of quantum Hall states \cite{PhysRevLett.107.116801,PhysRevX.7.041032,PhysRevLett.123.146801,PhysRevB.102.045145,PhysRevLett.126.076604} and bosonic phenomena such as superfluidity  and Bose-Einstein condensation \cite{peotta2015superfluidity,PhysRevLett.127.170404,torma2021superfluidity,guan2022re} in flat bands.

Flat, or nearly dispersionless, bands provide an ideal arena for strongly correlated states. The most prominent example thereof is the quantum Hall system exhibiting exactly flat bands in the continuum limit, and a rich phenomenology of strongly correlated states \cite{RevModPhys.89.025005}.  
Flat bands of lattice models are known to in principle exhibit an even richer phenomenology \cite{PhysRevB.48.8890,sheng2011fractional,PhysRevX.1.021014,PhysRevB.86.201101,PhysRevLett.109.186805,PhysRevB.86.241112,PhysRevB.86.241111,PhysRevB.87.205136,PhysRevB.87.205137,bergholtz2013topological}, for which the recently engineered superlattice Moir\'e materials provide remarkably versatile flat-band structures \cite{Bistritzer12233,andrei2020graphene,PhysRevLett.127.246403,wang2021hierarchy,PhysRevResearch.2.023237,PhysRevLett.122.106405,lado2021designer} that can be controlled in experiments \cite{balents2020superconductivity,xie2021fractional}. 

In this Letter, we show that the Fubini-Study (FS) metric \cite{provost1980riemannian,PhysRevLett.65.1697,kobayashi2009foundations} has a profound impact on the low-energy physics of strongly interacting lattice models and that it can induce novel fermionic phases in lattice flat-band systems that have no direct analogue in continuum Landau levels. We derive an emergent kinetic energy, which explicitly depends on the FS metric, through two distinct but mutually converging approaches: via a particle-hole (PH) transformation and a self-consistent Hartree-Fock calculation. 
While these results in principle have a wide range of applicability, we here focus on applying this to realistic Moir\'e systems for which we find that the quantum metric plays a preeminent role. Indeed, recent experiments have identified a large number of symmetry broken states \cite{xie2021fractional,pierceUnconventionalSequenceCorrelated2021,polshyn2022topological,bhowmik2022broken}. Here, we provide an intuitive picture of the symmetry breaking: electrons (holes) tend to occupy regions of Brillouin zone (BZ) with short (long) quantum distances as quantified by a small (large) quantum metric. The main difference between this work and previous ones \cite{PhysRevB.90.165139,PhysRevLett.114.236802,jackson2015geometric,PhysRevResearch.2.023237,PhysRevLett.127.246403} about the quantum metric is that we do not seek analogs between flat bands and Landau levels at the single-particle level. The role of the quantum metric here purely comes from many-body interacting effects \cite{PhysRevLett.111.126802}. The quantum distance turns out to be vital in reducing a strongly interacting question to a weakly interacting question. 

\begin{figure}
    \centering
    \includegraphics[width=\linewidth]{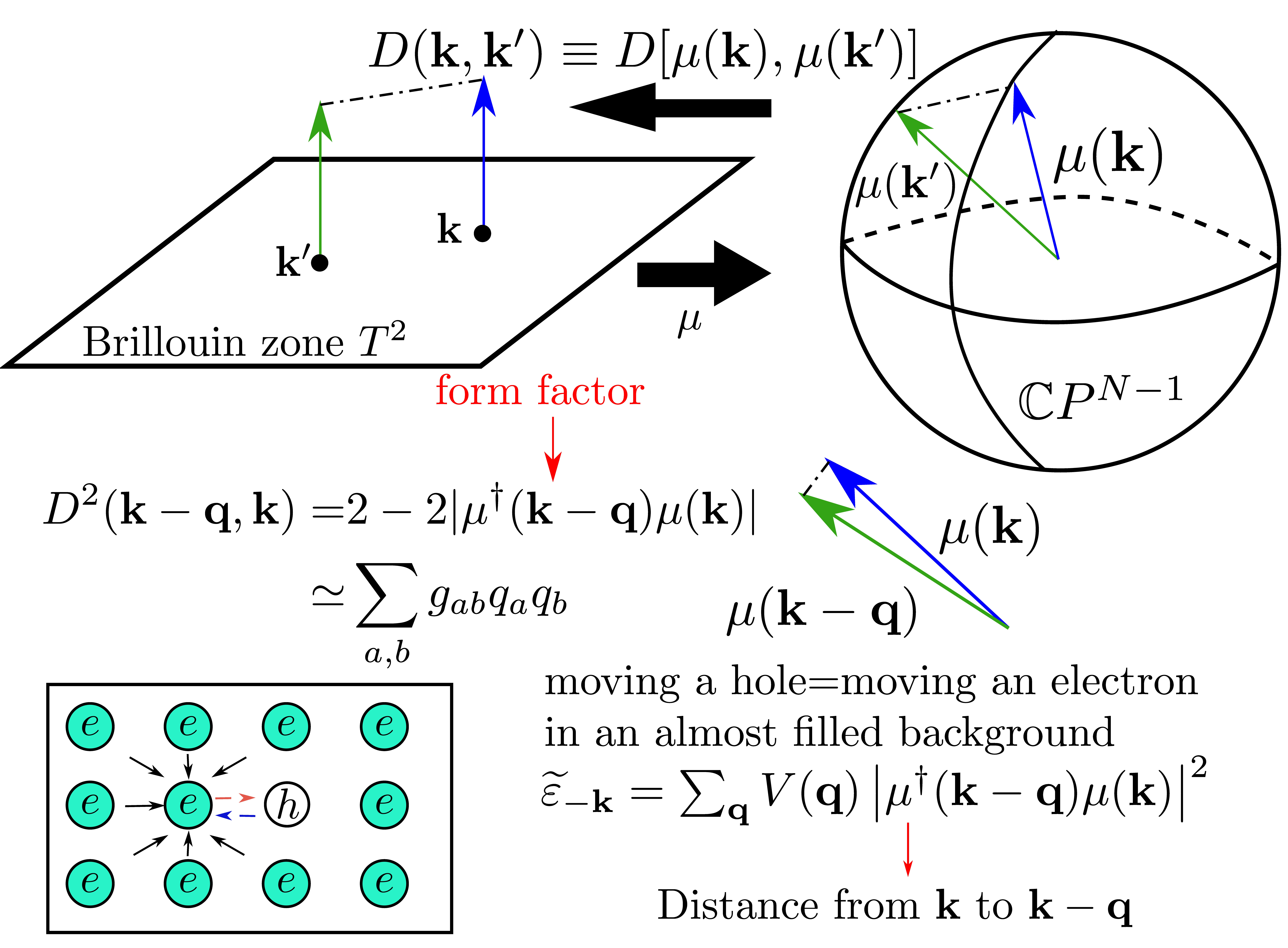}
    \caption{Geometric interpretation of the form factor in band projections. The Bloch state $\mu(\mathbf k)$ maps the Brillouin zone to a complex projective space $\mathbb{C}P^{N-1}$. The geometric information of the quantum states on $\mathbb{C}P^{N-1}$ can be pulled back to the Brillouin zone. The norm of the form factor directly reflects the quantum distance and the Fubini-Study metric. The exchange interaction $\tilde\varepsilon$ relies on the quantum distance and brings non-trivial hole dispersion.}
    \label{fig_illus}
\end{figure}

\paragraph{Quantum metric}- In quantum physics, there is a natural notion of distance between quantum states by regarding them as normalized complex vectors \cite{provost1980riemannian,PhysRevLett.65.1697,kobayashi2009foundations,CagliariDiFabioLandi+2015+1729+1742}. In a tight-binding lattice model, the periodic part $\mu(\mathbf k)$ of the Bloch states $\langle\mathbf x|\mathbf k\rangle=\exp(i\mathbf k\cdot\mathbf x)\mu(\mathbf k)$ is a vector of finite dimension. This gives rise to a distance measuring the difference between Bloch vectors at momenta $\mathbf k$ and $\mathbf k'$
\begin{equation}
D^2(\mathbf k,\mathbf k')=2-2|\mu^\dagger(\mathbf k)\mu(\mathbf k')|,\label{distance}
\end{equation}
as illustrated in Fig.~\ref{fig_illus}. Here $\mu^\dagger(\mathbf k)\mu(\mathbf k')$ is the inner product $\sum_{j=1}^N \mu^\ast_j(\mathbf k)\mu_j(\mathbf k')$ for a model with $N$ bands. To gain some intuition, note that identical Bloch vectors at $\mathbf k$ and $\mathbf k'$ yield $D=0$ while orthogonal ones give $D=\sqrt{2}$ consistent with Pythagoras' theorem.

At small separation $\mathbf q$, this distance leads to a FS metric $g_{ab}(\mathbf k)$: 
\begin{equation}D^2(\mathbf k-\mathbf q,\mathbf k)\approx \sum_{a,b=1}^2 g_{ab}(\mathbf k)q_aq_b.
\end{equation}
Equivalently, the FS metric can be expressed as \cite{PhysRevB.90.165139,PhysRevB.104.045104}
\begin{align}
    2g_{ab}(\mathbf k)=&\partial_{a}\mu^\dagger(\mathbf k)\partial_{b}\mu(\mathbf k)-[\partial_{a}\mu^\dagger(\mathbf k)\mu(\mathbf k)][\mu^\dagger(\mathbf k)\partial_{b}\mu(\mathbf k)]\nonumber\\
    &+(a\leftrightarrow b).
\end{align} 
One may note that the distance (\ref{distance}) is different from the geodesic distance \cite{PhysRevLett.65.1697} or the Hilbert–Schmidt quantum distance \cite{PhysRevA.54.1844,rhim2020quantum}, but all these definitions coincide when the quantum distance is small leading to the same effective metric. Mathematically the terminology of distance and metric tensor \cite{howes2012modern} introduced here may need a careful treatment. More details can be found in the supplementary material (SM) \cite{suppm}.

\paragraph{Emergent kinetic energy}- In a flat-band model the relevant physics is given by the projected interaction of electrons:
\begin{align}
    H=&\frac{1}{2}\sum_{\mathbf q ,\mathbf k,\mathbf k'}V(\mathbf q)\left[\mu^\dagger(\mathbf k-\mathbf q) \mu(\mathbf k)\right] \left[\mu^\dagger(\mathbf k'+\mathbf q) \mu(\mathbf k')\right] \nonumber\\
    &\times c^\dagger_{\mathbf {k-q}}c^\dagger_{\mathbf {k'+q}} c_{\mathbf k'} c_{\mathbf k},
\end{align}
The key ingredient is the projected density operator at momentum $\mathbf q$, which is the product of the operators $c^\dagger_{\mathbf k-\mathbf q} c_{\mathbf k}$ and the form factor $\mu^\dagger(\mathbf k-\mathbf q) \mu(\mathbf k)$. We consider a single band. i.e, we freeze any additional degrees of freedom such as the spin and the valley. As there is no kinetic energy, the electrons are strongly interacting and the many-body state seems to be very complicated. Previous work \cite{PhysRevLett.111.126802,PhysRevLett.124.106803} notices that under a PH transformation the Fermi liquid may be a good candidate for ground states. At a first glance, however, the Hamiltonian seems PH symmetric. The enigma lies in the fluctuating band geometry and the correlated nature of hole-like degrees of freedom. A PH transformation maps $c_{\mathbf k}\to d^\dagger_{-\mathbf k}$.  In addition to the transformation on operators, we also need to transform the concomitant reference state, from the vacuum of electrons $|\Omega\rangle$ to the vacuum of holes $|\Omega'\rangle=\prod_{\mathbf k}c^\dagger_{\mathbf k}|\Omega\rangle$. The low-energy excitations of $|\Omega'\rangle$ is described by creating a few holes. To better capture this physical process, we need to normal order the Hamiltonian by moving all hole creation operators to the right of hole annihilation operators. This procedure gives the following Hamiltonian
\begin{align}
     \bar{\bar H}=&\sum_{\mathbf k}\varepsilon_{\mathbf k}d^\dagger_{\mathbf k}d_{\mathbf k}+\frac{1}{2}\sum_{\mathbf q,\mathbf k,\mathbf k'}  V(\mathbf q) \left[\mu^\dagger(-\mathbf k) \mu(\mathbf q-\mathbf k)\right]\nonumber\\
     &\times \left[\mu^\dagger(-\mathbf k') \mu(-\mathbf k'-\mathbf q)\right] d^\dagger_{\mathbf k-\mathbf q}d^\dagger_{\mathbf k'+\mathbf{q}} d_{\mathbf k'} d_{\mathbf k}.
     \label{eq_phtransformH}
\end{align}
Apart from a similar interaction term to electrons, holes receive a quadratic term that is generically dispersing (an exception is Landau levels where it is constant) \cite{PhysRevLett.111.126802}. This term reflects the fact that the kinetics of a hole is described by how an electron moves in an electron background (see Fig.~\ref{fig_illus}). It can be decomposed into two parts $\varepsilon= \overline\varepsilon +\widetilde \varepsilon$. The first constant term comes from a uniform background repulsion and is equivalent to a chemical potential (details in the SM~\cite{suppm}). The second term is nontrivial, resulting from the exchange interaction: $\widetilde\varepsilon_{-\mathbf k}=\sum_{\mathbf q} V(\mathbf q)\vert\mu^\dagger(\mathbf k-\mathbf q) \mu(\mathbf k)\vert^2$ where the form factor norm enters.

A key observation is that the form factor norm occurring in $\widetilde\varepsilon_{-\mathbf k}$ describes the distance between two Bloch states $\mu(\mathbf k)$ and $\mu(\mathbf k-\mathbf q)$. The exchange interaction is thus jointly determined by the \emph{interaction potential} and the \emph{quantum distance}. The kinetic energy of holes is dispersing as long as the quantum geometry pulled back to the BZ is not uniform. For small momentum $\mathbf q$, $\widetilde\varepsilon_{-\mathbf k}$ can be expanded by the FS metric $\vert\mu^\dagger(\mathbf k-\mathbf q) \mu(\mathbf k)\vert^2\simeq 1- \sum_{ab}q_a q_bg_{ab}(\mathbf k)$:
\begin{align}
    \widetilde\varepsilon_{-\mathbf k}
    \simeq\sum_{\mathbf q} V(\mathbf q)e^{-\sum_{ab} q_a q_bg_{ab}(\mathbf k)},\label{eq_snmetric}
\end{align}
where we further approximate the parabolic expansion by lifting it to the exponent so that the norm is still positive at large $\mathbf q$. This ansatz of form factors also appears in the context of ideal fractional Chern insulators (FCI) \cite{PhysRevB.90.165139}, although the focus here is on the opposite aspects, namely the qualitative deviations from Landau level physics.

\begin{figure}
    \centering  
    \includegraphics[width=1.\linewidth]{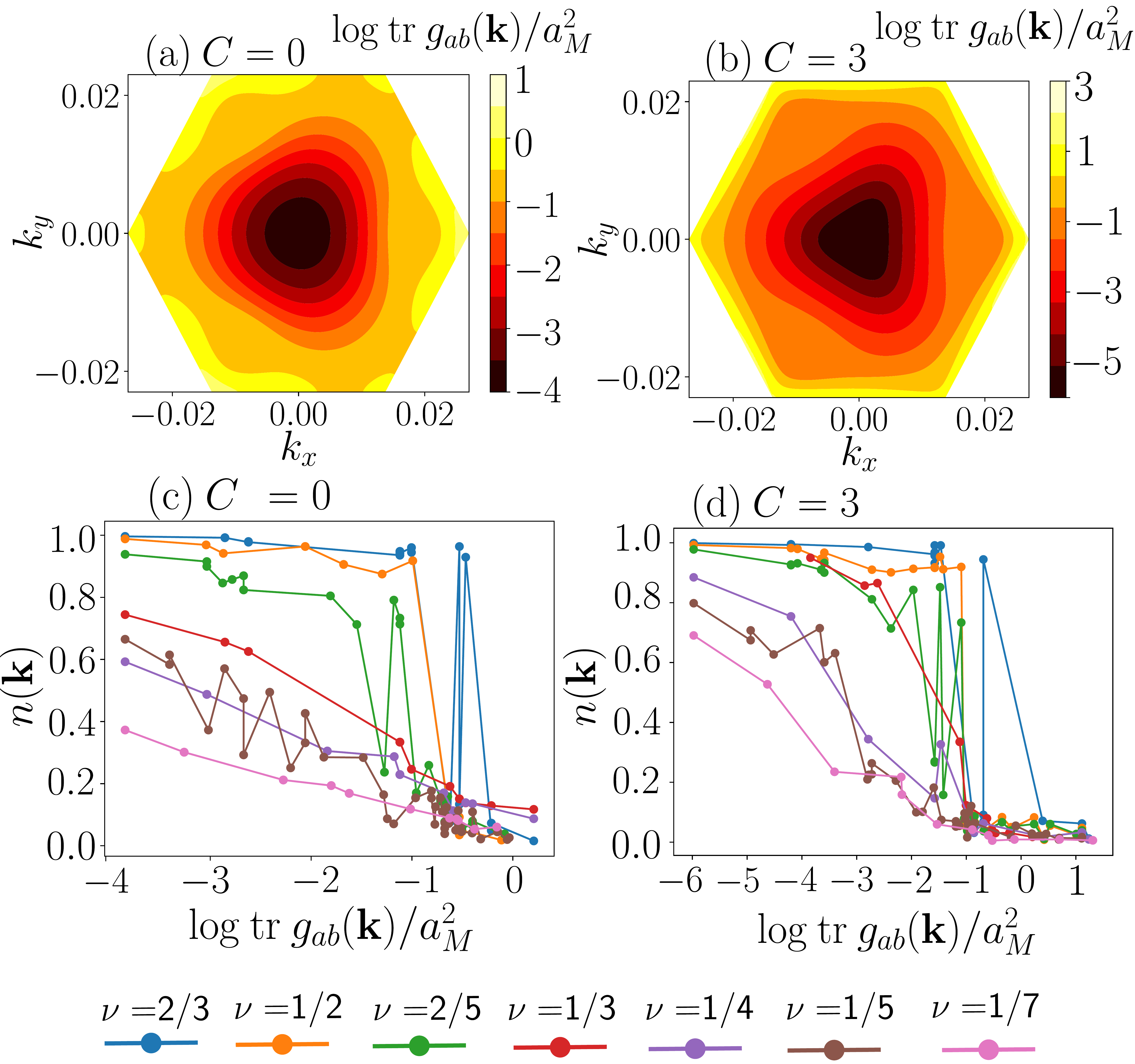}
    \caption{Metric-induced phases in TLG-hBN $C=0$ valence band and $C = 3$ valence band obtained by switching the sign of the gate-voltage: (a)-(b) Contour plots of $\textrm{log tr }g_{ab}(\mathbf{k})/a_M^2$ where $a_M$ is the Moir\'e unit cell lattice constant.  (c)-(d) Lowest energy ground state electron occupation $n(\mathbf{k}) = \langle c^\dagger_{\mathbf{k}} c_{\mathbf{k}} \rangle$ obtained from exact diagonalization \cite{suppm} vs  $\textrm{log tr }g_{ab}(\mathbf{k})/a_M^2$ for as a function of electron band filling $\nu$. We use gate voltage $U = \pm 0.02$ eV. \cite{chittariGateTunableTopologicalFlat2019a}}
    \label{fig_TLG-hBN}
\end{figure}

Intuitively, one may expect that the hole dispersion and the Fermi-liquid idea would only be useful at high electronic band filling. Remarkably, however, the Fermi liquid prevails in a wide range of fillings $\nu$ \cite{PhysRevLett.124.106803} for Moir\'e systems. This can be understood from a self-consistent picture of Landau's Fermi liquids, yielding a Fock energy induced by the interaction \cite{suppm}
\begin{equation}
     E_{\mathbf k}\simeq-\sum_{\mathbf q}V(\mathbf q)e^{-\sum_{ab} q_a q_bg_{ab}(\mathbf k)}\langle c^\dagger_{\mathbf k-\mathbf q}c_{\mathbf k-\mathbf q}\rangle\label{eq_scne}.
\end{equation}
Because of the non-uniform geometry, electrons prefer to fill the areas in the BZ with small $\textrm{tr }g$ (Fig.~\ref{fig_TLG-hBN}), where they can benefit from a lower Fock energy due to the smaller quantum distance. This renders the Fermi liquid a natural ground-state candidate also in the electron picture. The hole dispersion can be regarded as the extreme case when the band is completely filled ($\langle c^\dagger_{\mathbf k}c_{\mathbf k}\rangle=1$ for all $\mathbf k$). As we will show, Moir\'e systems have rapidly-decaying form factors. Then the partial summation \eqref{eq_scne} over the electron-filled region is not very different from the whole BZ summation \eqref{eq_snmetric}. So the hole energy gives a good estimate of the self-consistent Hartree-Fock Hamiltonian of the electrons, $ E_{\mathbf k}\simeq -\tilde\varepsilon_{-\mathbf k}$, where the minus signs come from the PH relation. Explicit comparisons can be found in the SM \cite{suppm}.

Equations~\eqref{eq_snmetric}, \eqref{eq_scne} and their relevance to Moir\'e materials are the main results of this Letter. The non-uniform quantum geometry and the finite electron density induce a non-trivial emergent kinetic energy in flat bands. Locally, such a non-uniform geometry is captured by the FS metric and the kinetic energy can be obtained by integrating the interaction potential with the metric. This emergent kinetic energy naturally appears in the hole Hamiltonian as hole degrees of freedom have the completely filled band as a reference state. In contrast, Landau levels have an exactly flat geometric structure and the Hamiltonian (any band projected translation invariant two-body interaction) is always PH symmetric.

\paragraph{Metric-dominated conditions}- We expect that Eq.~\eqref{eq_snmetric} gives the dominant contribution to the emergent kinetic energy when the product of the interaction potential and the form factor decays fast enough in $\mathbf q$. The form factor reflects the quantum distance. As the BZ is fixed to be 2D, we can envisage its image $\mu(\mathbf k)$ to be some 2D subspace of the complex projective space $\mathbb{C}P^{N-1}$. The average distance between $\mu(\mathbf k-\mathbf q)$ and $\mu(\mathbf k)$ thus depends on to which extent this subspace extends in $\mathbb{C}P^{N-1}$. A natural condition is that $N$ should be large, otherwise most states in $\mathbb{C}P^{N-1}$ are nearby. Especially for a topologically non-trivial band, the image of the BZ tends to span much of $\mathbb{C}P^{N-1}$. Thus a system with sufficient many bands is likely to support a fast-decaying form factor. An excellent approximation is obtained by considering the finite thickness of the 2D material. This results in a potential decaying in the momentum space, for example, the Zhang-Das Sarma potential $V(\mathbf q)=2\pi e^2\exp(-lq)/q$, where $l$ is the sample thickness \cite{PhysRevB.33.2903}. 

\paragraph{Applications to Moir\'e materials}- The number of minibands in Moir\'e systems is estimated as $N\simeq (|\mathbf Q_0|/|\mathbf G_0|)^2\sim 10^3$, where $\mathbf G_0$ is the primitive Moir\'e reciprocal lattice vector and $\mathbf Q_0$ is the primitive reciprocal lattice vector of the original lattice. So Moir\'e materials satisfy the condition of sufficiently many bands. In continuum approximations, the Moir\'e Bloch states $|\psi_{\mathbf k}\rangle$ are constructed by superposing the original Bloch states $|\mathbf k+\mathbf G\rangle$ differed by Moir\'e reciprocal wave vectors, $|\psi_{\mathbf k}\rangle=\sum_{\mathbf G}\mu(\mathbf k,\mathbf G)|\mathbf k+\mathbf G\rangle$. Here we omit other indices for spins or valleys. The form factor is obtained as
$\sum_{\mathbf G} \mu^\ast(\mathbf k-\mathbf q,\mathbf G)\mu(\mathbf k,\mathbf G)$. A natural definition of Moir\'e Bloch vectors is given by $\mu(\mathbf k,\mathbf G)$ with $\mathbf G$ as an index of components. We show in the SM \cite{suppm} that the Moir\'e form factors are indeed quickly decaying.

Different from conventional tight-binding models, the norm of the Moir\'e form factor is not periodic in $\mathbf q$. A usual Bloch vector $\mu(\mathbf k)$ and its BZ translated counterpart $\mu(\mathbf k+\mathbf Q)$ only differs by a phase $\phi_{\mathbf k,\mathbf Q}$. In contrast, the Moir\'e BZ translated pairs are different by an additional cyclic transformation on its components $\mu(\mathbf k+\mathbf G',\mathbf G)=\exp(i\phi_{\mathbf k,\mathbf G'})\mu(\mathbf k,\mathbf G+\mathbf G')$. For this reason, even the contribution $\overline\varepsilon_{\mathbf k}$ is (weakly) dispersing and $\widetilde\varepsilon_{\mathbf k}$ becomes more complicated \cite{suppm}. However, all of these differences are happening at momenta equal or larger than $\mathbf G_0$ and since the Moir\'e form factors are quickly decaying, these modifications are minimal. Hence $\eqref{eq_snmetric}$ remains accurate.

\begin{figure}
    \centering
    \includegraphics[width=\linewidth]{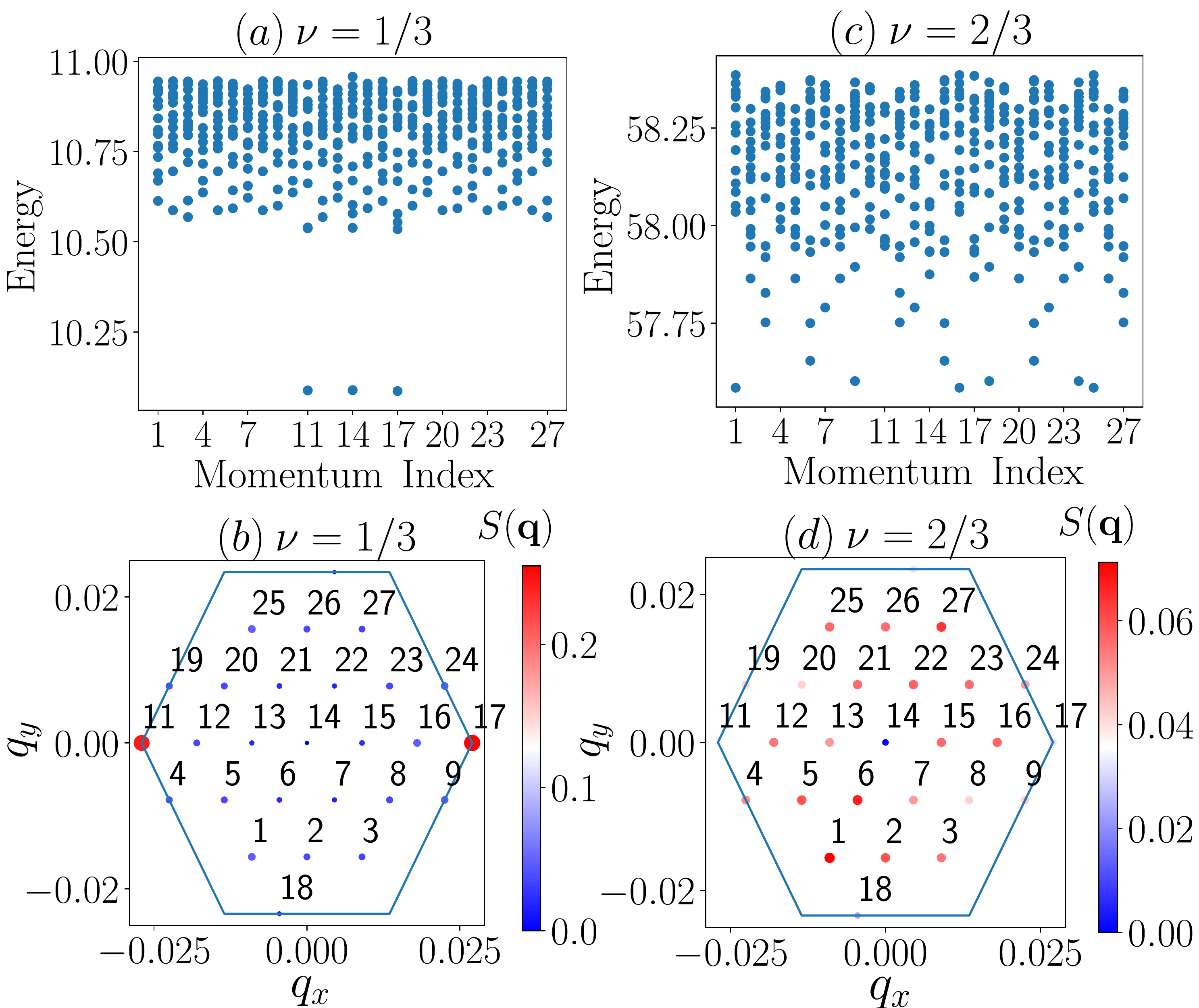}
    \caption{Particle-hole asymmetry in TLG-hBN $C = 0$ valence band: (a)-(b) Evidence of charge density wave at $\nu = 1/3$. (a) Many-body spectrum showing 3-fold degeneracy at 3 different momenta corresponding to the 3 possible charge configurations when the unit-cell is tripled. (b) The projected static structure constant calculated in the lowest energy ground state $S(\mathbf{q}) = \langle \hat{\rho}^{\rm proj}_{\mathbf{q}} \hat{\rho}^{\rm proj}_{-\mathbf{q}} \rangle $  with $\hat{\rho}^{\rm proj}_{\mathbf{q}} = \sum_{\mathbf{k}} \mu^\dagger(\mathbf{k} - \mathbf{q}) \mu(\mathbf{k}) c^\dagger_{\mathbf{k}-\mathbf{q}} c_{\mathbf{k}}$. Prominent peaks are observed at the $\mathbf{K}$ points. (c)-(d) Absence of charge density wave at $\nu = 2/3$. (c) Many-body spectrum with no clear separatred low energy sector. (d) Featureless static structure factor $S(\mathbf{q})$ with no prominent peaks. 
    }
    \label{fig_cdw}
\end{figure}

\begin{figure*}
    \centering
    \includegraphics[width=1.0\linewidth]{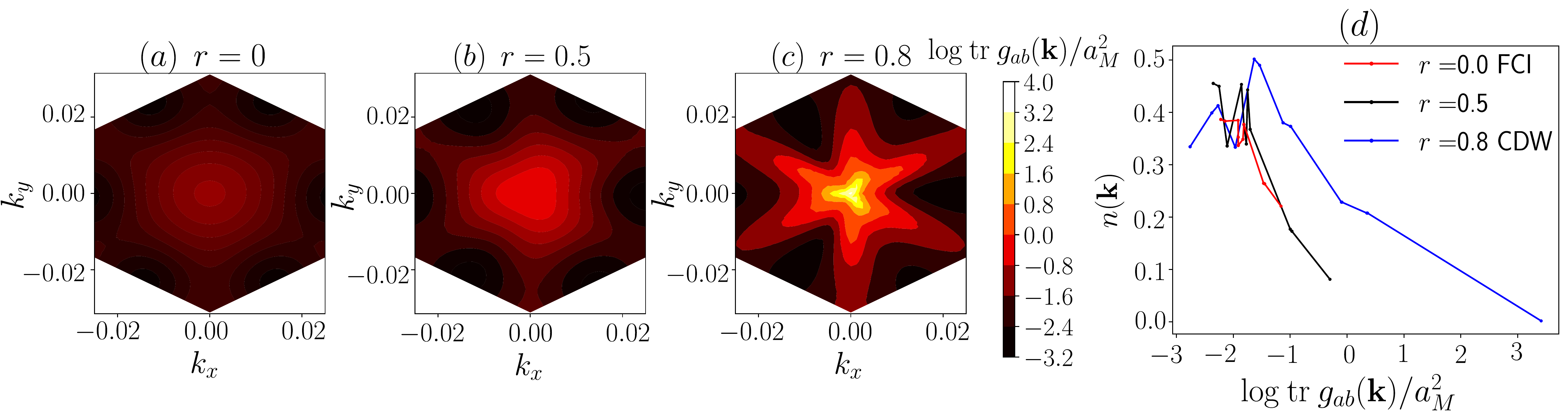}
    \caption{Fubini-Study metric effect in the spin and valley polarized $C = 1$ valence band in TBG-hBN: (a)-(c) Contour plots of $\textrm{log tr }g_{ab}(\mathbf{k})/a_M^2$ at twist angle $\theta = 1.05^\circ$ and mass term \cite{zhangTwistedBilayerGraphene2019} $M = 15$ meV for  different values of $r = w_0/w_1$. $a_M$ is the Moir\'e unit cell lattice constant  (d) Lowest energy ground state electron occupation $n(\mathbf{k}) = \langle c^\dagger_{\mathbf{k}} c_{\mathbf{k}} \rangle$  obtained from exact diagonalization \cite{suppm} at $\nu = 1/3$ vs  $\textrm{log tr }g_{ab}(\mathbf{k})/a_M^2$ for different values of $r$. The criteria for FCI agree with counting rules in Ref.~\onlinecite{bernevigEmergentManybodyTranslational2012}. }
    \label{fig_fcivscdw}
\end{figure*}

Equipped with the connection between the FS metric and an emergent dispersion, we proceed to analyze possible phases in prominent Moir\'e systems such as trilayer graphene and twisted bilayer graphene, both aligned with hexagonal boron nitride. Both these two setups and several other related ones have attracted ample interest as tunable platforms for studying strongly correlated phases such as correlated insulators and superconductivity \cite{caoCorrelatedInsulatorBehaviour2018,caoUnconventionalSuperconductivityMagicangle2018,chenEvidenceGatetunableMott2019,chenSignaturesTunableSuperconductivity2019,chenTunableCorrelatedChern2020,chenTunableOrbitalFerromagnetism2022,choiCorrelationdrivenTopologicalPhases2021,choiInteractiondrivenBandFlattening2021,linSpinorbitDrivenFerromagnetism2022,luSuperconductorsOrbitalMagnets2019,nuckollsStronglyCorrelatedChern2020a,ohEvidenceUnconventionalSuperconductivity2021,pierceUnconventionalSequenceCorrelated2021,saitoHofstadterSubbandFerromagnetism2021,sharpeEmergentFerromagnetismThreequarters2019,sharpeEvidenceOrbitalFerromagnetism2021,tsengAnomalousHallEffect2022,wuChernInsulatorsVan2021,xieSpectroscopicSignaturesManybody2019,yankowitzTuningSuperconductivityTwisted2019,zhouHalfQuartermetalsRhombohedral2021}.

\paragraph{Trilayer graphene aligned with boron-nitride (TLG-hBN)} - The Moir\'e system trilayer graphene aligned with hBN hosts a nearly flat valence band below charge neutrality upon applying a gate voltage across the layers \cite{chittariGateTunableTopologicalFlat2019a,zhangNearlyFlatChern2019}. At the non-interacting level neglecting possible strain effects \cite{gonzalezTopologicalPhasesLayer2021}, the flat band could be topologically trivial with Chern number $C = 0$ or topologically non-trivial with $C = 3$ depending on the sign of the applied gate voltage.  It was theoretically shown \cite{PhysRevLett.124.106803} that this system could host Fermi liquid phases dictated by the single-hole dispersion for a wide range of filling. Remarkably, the emergent dispersion greatly influences the average electron occupation down to very low fillings. Regardless of the value of the Chern number, TLG-hBN valence band has a non-trivial and far from flat FS metric quantified by the trace $\textrm{tr }g(\mathbf{k})$ shown in Fig. \ref{fig_TLG-hBN}(a) and Fig. \ref{fig_TLG-hBN}(b). We find the electron occupation in momentum space $n(\mathbf{k}) $ to correlate well with $\textrm{log tr }g(\mathbf{k})/a_M^2$
 as indicated in Figs.  \ref{fig_TLG-hBN}(c)-(d) for representative filling fractions $\nu \leq 2/3$. This shows that the origin of such correlation is mainly induced by the highly fluctuating FS metric that results in an kinetic energy with large bandwidth. As shown in Fig. \ref{fig_TLG-hBN}(a) and Fig. \ref{fig_TLG-hBN}(b), we find $\textrm{log tr }g(\mathbf{k})/a_M^2$ to share the same qualitative features with the hole dispersion \cite{PhysRevLett.124.106803} up to very high electron fillings where it starts to deviate from the hole energy. 
The gradual disappearance of well-defined Fermi surfaces, manifested in the sharp jumps in the occupation $n(\mathbf{k})$ as shown in Figs. \ref{fig_TLG-hBN}(c)-(d), as the electron filling $\nu$ decreases signals possible transitions from the Fermi liquid state to competing states. Charge density waves (CDW) are natural candidates at fillings that are commensurate with the triangular Moir\'e lattice. To illustrate this, we provide numerical evidence for a possible CDW at $\nu = 1/3$ for the $C = 0$ band through the 3-fold ground state degeneracy shown in Fig. \ref{fig_cdw}(a) and the static structure factor peaks shown in Fig. \ref{fig_cdw}(b). 
The effects of quantum geometry are however present even when there is an absence of a well defined Fermi surface. Although the sharp Fermi surface is blurred (c.f $\nu = 1/3$ in Fig. \ref{fig_TLG-hBN}.(c)), the ground state occupation correlates well with  $\textrm{tr }g$ as the electrons tend to stay in regions with small $\textrm{tr }g$. Moreover, the non-uniform quantum metric reflects a strong PH asymmetry; we find no evidence of CDW at the PH dual filling $\nu = 2/3$ as illustrated in Figs. \ref{fig_cdw}(c)-(d). Indeed, the occupation $n(\mathbf{k})$ (c.f $\nu = 2/3$ in Fig. \ref{fig_TLG-hBN} (c)), when averaged over a number of low lying energy states, shows clear signs of a Fermi liquid state \cite{suppm}. This asymmetry may be attributed to the different Fermi-surface geometries corresponding to $\nu=1/3$ and $\nu=2/3$, dictated by the FS metric. The systematic identification of the ground states at different commensurate fillings and the nature of transitions provides avenues for future work starting from the insights presented here, including the study of possible instabilities starting from Fermi surfaces suggested by the FS metric. 


\paragraph{Twisted bilayer graphene aligned with boron-nitride (TBG-hBN)}- Next we study twisted bilayer graphene aligned with hBN. The alignment with hBN gaps out the flat bands of twisted bilayer graphene and make them acquire a non-zero Chern number $C = \pm 1$ \cite{zhangBridgingHubbardModel2019,zhangTwistedBilayerGraphene2019}. Earlier theoretical studies \cite{PhysRevLett.124.106803,repellinChernBandsTwisted2020,PhysRevResearch.2.023237} have predicted the possibility of realizing zero magnetic field FCI state at fractional fillings of the flat bands of TBG-hBN. These FCI states compete with possible CDWs at commensurate fillings \cite{PhysRevB.103.125406}. Such competition depends on the relaxation ratio $r = w_0/w_1$ where $w_0$ and $w_1$ are the interlayer tunneling strengths at the AA stacked and AB stacked regions respectively. Assuming spin and valley polarization, we focus on one the valence band at filling fraction $\nu = 1/3$ and vary $r$. (Qualitatively similar results were obtained in the valence band at $\nu = 2/3$ and in the conduction band at both $\nu = 1/3$  and $\nu = 2/3$.) As the value of $r$ increases, FS metric fluctuations become more prominent and a transition from FCI to CDW occurs (Fig. \ref{fig_fcivscdw}). The critical value of $r$ of such transition is sensitive to the model parameters such as $w_1$ and the twist angle $\theta$ \cite{parker2021field}. 
Recent experiments \cite{xie2021fractional} have confirmed the existence of FCI states in TBG, albeit with the application of a weak magnetic field. It was argued \cite{xie2021fractional,parker2021field} that the role of the weak magnetic field, similar to a smaller $r$, is to make the Berry curvature more uniform and thus stabilizing the zero-field FCI.
Noting that in this system, the FS metric fluctuates in sync with the Berry curvature \cite{PhysRevResearch.2.023237,PhysRevLett.127.246403,parker2021field}, the magnetic field or the ratio $r$ also flattens the FS metric. As a consequence, we here provide a complementary interpretation of the competition between FCI and CDW based on the FS metric. FCIs prefer the electron density to be uniform in momentum space and the Berry curvature, the effective ``magnetic field'', to be non-vanishing in these regions. If the metric is uniform, electrons do not have priorities in the BZ and this condition can be satisfied. As the ratio $r$ increases, however, the FS metric becomes less uniform as shown Fig. \ref{fig_fcivscdw}(a)-(c). Accordingly, there is a tendency for the electrons to occupy states with lower $\textrm{tr }g(\mathbf{k})$ to minimize the Fock energy according to equations \eqref{eq_phtransformH}-\eqref{eq_scne}. This leads to a varying $n(\mathbf{k})$, destabilizing the FCI state being is a liquid of roughly uniform density. This is highlighted in Fig. \ref{fig_fcivscdw}(d) where we show how the occupation $n(\mathbf{k})$ evolves for different values of $r$ as a function of $\textrm{log tr }g(\mathbf{k})/a_M^2$.  

\paragraph{Discussion}- We have shown that the quantum geometry in terms of the Fubini-Study metric explicitly appears in the effective Hamiltonian description of strongly interacting electrons in a flat band, yielding an emergent kinetic energy stemming from the collective interactions. This suggests generic and physically intuitive picture of symmetry breaking states--and their microscopic provenance--which have been observed in recent experiments on Moir\'e materials \cite{xie2021fractional,pierceUnconventionalSequenceCorrelated2021,polshyn2022topological,bhowmik2022broken}.

Our theory also provides an alternative criterion for the stability of FCIs. Whereas previous studies \cite{PhysRevB.85.241308,PhysRevB.90.165139,PhysRevB.105.045144} were based on the similarity of the density algebra to quantum Hall effects, here we gives a microscopic interpretation of band geometry. The uniformity of the Fubini-Study metric dictates where the electrons are likely to condensate in momentum space. The observed FCI/CDW transitions in Ref.~\onlinecite{xie2021fractional} may also be explained by the more uniform metric tuned through the magnetic field or interlayer interaction ratio $w_0/w_1$. As a comparison, the interpretation of the Berry curvature from a microscopic view is less clear \cite{varjas2021topological}. 

The paramount role of the Fubini-Study metric in fractionally filled Moir\'e bands may serve as a guiding principle in materials design aiming to realise various exotic phases. Moreover, it effectively turns experimentally relevant instances of an extremely strongly interacting problem into an effectively weakly interacting one, thus enabling the use of standard techniques of theoretical physics.
\\
\acknowledgments{We acknowledge helpful discussions with Yonglong Xie, Shaowen Chen, Zhao Liu, Ipsita Mandal, and Daniel Varjas. The authors are supported by the Swedish Research Council (VR, grant 2018-00313), the Wallenberg Academy Fellows program of the Knut and Alice Wallenberg Foundation (2018.0460) and the G\"oran Gustafsson Foundation for Research in Natural Sciences and Medicine.}

\bibliography{main}


\begin{widetext}

\section{Supplemental Material}
\renewcommand{\theequation}{S\arabic{equation}}
\setcounter{equation}{0}
\renewcommand{\thefigure}{S\arabic{figure}}
\setcounter{figure}{0}
\renewcommand{\thetable}{S\arabic{table}}
\setcounter{table}{0}
\section{Quantum distance and metric}
The Bloch state $\langle\mathbf x|\mathbf k\rangle=\exp(i\mathbf k\cdot\mathbf x)\mu(\mathbf k)$ on an $N$-band 2D lattice defines a map from the Brillouin zone (BZ) to the complex projective space $\mathbb{C}P^{N-1}$. This is represented by the normalized $N$-component vector $\mu(\mathbf k)$, describing the wave functions on the $N$ orbitals inside a unit cell. The $\mathbb{C}P^{N-1}$ space is equipped with a natural geometric structure reflected in the quotient relation $\mathbb{C}P^{N-1}\simeq S^{2N-1}/U(1)$. The sphere $S^{2N-1}$ is metrizable as a subspace of $\mathbb{R}^{2N}$ and its distance function is invariant under the action of $U(1)$. So there is a natural distance for the quantum states, given by $D^2[\mu,\mu']=2-2|\mu^\dagger\mu'|$. From this distance, we can find a Fubini-Study (FS) metric in the tangent space of $\mathbb{C}P^{N-1}$ \cite{kobayashi2009foundations,provost1980riemannian,PhysRevLett.65.1697}. By pulling back such Riemann structure using the map $\mu$, a metric tensor can be constructed on the BZ \cite{PhysRevB.90.165139,PhysRevB.104.045104}
\begin{align}
    2g_{ab}(\mathbf k)=&\partial_{a}\mu^\dagger(\mathbf k)\partial_{b}\mu(\mathbf k)-[\partial_{a}\mu^\dagger(\mathbf k)\mu(\mathbf k)][\mu^\dagger(\mathbf k)\partial_{b}\mu(\mathbf k)]\nonumber\\
    &+(a\leftrightarrow b),
\end{align}
where $\mu^\dagger\mu$ is the inner product $\sum_j \mu^\ast_j\mu_j$. This is the FS metric defined in the BZ and measures the infinitesimal distance between Bloch states at different positions of the BZ: $g_{ab}$ is the leading expansion coefficient of the distance function given by the pullback $D^2(\mathbf k,\mathbf k')\equiv D^2[\mu(\mathbf k),\mu(\mathbf k')]=2-2|\mu^\dagger(\mathbf k)\mu(\mathbf k')|$ \cite{provost1980riemannian,CagliariDiFabioLandi+2015+1729+1742}. To be more mathematically cautious, the pullback $D(\mathbf k,\mathbf k')$ should be termed as a pseudometric \cite{howes2012modern}. As different $\mathbf k$ and $\mathbf k'$ may be mapped to the same vector $\mu$, $D(\mathbf k,\mathbf k')=0$ can hold for $\mathbf k\ne \mathbf k'$. The FS metric $g_{ab}(\mathbf k)$ on the BZ is not a Riemann metric. Since the tangent map $TBZ\to T \mathbb{C}P^{N-1}$ is not always injective, the BZ pullback $g_{ab}(\mathbf k)$ of the FS metric on $\mathbb{C}P^{N-1}$ is not guaranteed to be positive definite for all $\mathbf k$. In fact, one can show that the FS metric must be degenerate at some momentum of the BZ for two-band models \cite{PhysRevB.104.045104,varjas2021topological}.

\section{Projected density operators and form factors}

The interaction is expressed as the product of the potential with the density operators 
\begin{equation}
    \hat V=\frac{1}{2}\sum_{\mathbf q}V(\mathbf q):\hat\rho(\mathbf q)\hat\rho(-\mathbf q): ,
\end{equation}
where $\hat\rho(\mathbf q)$ is the Fourier transformation of the density operator in coordinate space $\hat\rho(\mathbf q)=\sum_{\mathbf x}\exp(-i\mathbf q\cdot\mathbf x)\hat\rho(\mathbf x)$. Here
we perform a normal ordering, moving all creation operators of electrons to the left of annihilation operators, so as to exclude the situation where an electron interacting with itself. In order to perform the band projection, it is better to rewrite it in the basis of Bloch states. This is more easily done with the help of its single-particle representation $\hat\rho(\mathbf q)=\exp(-i\mathbf q\cdot\hat{\mathbf x})$, where $\hat{\mathbf x}$ is the coordinate operator. In a lattice system, the Bloch states in the $n$-th band are expressed as $\langle\mathbf x|\mathbf k,n\rangle=\exp(i\mathbf k\cdot\mathbf x)\mu^{(n)}(\mathbf k) $, where $\mu^{(n)}$ is a vector with components labelling the orbitals. Here for simplicity we take all the orbitals to be exactly on the lattice site. The matrix element $\langle\mathbf k,n|\hat\rho(\mathbf q)|\mathbf k',n'\rangle$ is only non-vanishing for $\mathbf k=\mathbf k'-\mathbf q$. With these formulae, we found that the interaction takes the following second-quantized form
\begin{align}
    \hat V=\frac{1}{2}\sum_{\mathbf q,\mathbf k_i,n_i,n'_i}V(\mathbf q)\langle\mathbf k_1-\mathbf q,n_1|\hat\rho(\mathbf q)|\mathbf k_1,n'_1\rangle\langle\mathbf k_2+\mathbf q,n_2|\hat\rho(-\mathbf q)|\mathbf k_2,n'_2\rangle c^\dagger_{\mathbf k_1-\mathbf q,n_1}c^\dagger_{\mathbf k_2+\mathbf q,n_2}c_{\mathbf k_2,n'_2}c_{\mathbf k_1,n'_1},\label{eq_fint}
\end{align}
The above summation of momenta needs some caution. We must sum the momentum in the interaction potential over $\mathbf q\in \mathbb{R}^2$, since the interaction potential is not lattice-periodic. In contrast, the summations of $\mathbf k_1$ and $\mathbf k_2$ are limited to the first Brillouin zone in order to avoid multiple counting. Now assume that we are interested in the $m$-th band whose lattice kinetic energy is flat. The most relevant physics is controlled electrons in this band. The projection of the interaction is done by only keeping those terms $n_i=n'_i=m$ in Eq.~\eqref{eq_fint}
\begin{align}
    \hat V^{(m)}=&\frac{1}{2}\sum_{\mathbf q,\mathbf k,\mathbf k'}V(\mathbf q)\rho(\mathbf k,\mathbf q)\rho(\mathbf k',-\mathbf q) c^\dagger_{\mathbf k-\mathbf q,m}c^\dagger_{\mathbf k'+\mathbf q,m}c_{\mathbf k',m}c_{\mathbf k,m}\label{eq_smpjH},\\
    \rho(\mathbf k,\mathbf q)=&\langle\mathbf k-\mathbf q,m|\hat\rho(\mathbf q)|\mathbf k,m\rangle=\sum_j\mu^{(m)\ast}_j(\mathbf k-\mathbf q)\mu^{(m)}_j(\mathbf k).
\end{align}
The second line is the matrix elements of the projected density operator in the main text. The physics in the $m$-th band is determined by the interaction potential $V(\mathbf q)$ and the form factor $\rho(\mathbf k,\mathbf q)$ of the projected density. In later calculations, we will drop the index $m$ as done in the main text since it always refers to the flat band studied.  The Bloch states connected by reciprocal lattice vectors are related by
\begin{equation}
    |\mathbf k+\mathbf Q\rangle=e^{i\phi_{\mathbf k,\mathbf Q}}|\mathbf k\rangle,\quad \mu(\mathbf k+\mathbf Q)=e^{i\phi_{\mathbf k,\mathbf Q}}\mu(\mathbf k),\quad c^\dagger_{\mathbf k+\mathbf Q}=e^{i\phi_{\mathbf k,\mathbf Q}}c^\dagger_{\mathbf k},\label{eq_eqBZtrQ}
\end{equation}
where $\phi_{\mathbf k,\mathbf Q}$ is a phase. This phase can be chosen to be zero for topologically trivial bands but non-vanishing for topological Chern bands.

In the presence of superlattice hopping, Moir\'e Bloch states are given by $|\psi_{m,\mathbf k}\rangle=\sum_{\mathbf G,\sigma}|\mathbf k+\mathbf G,\sigma\rangle \mu_{\sigma}(m,\mathbf k,\mathbf G)$, where $\mathbf G$ is the Moir\'e reciprocal lattice vector and $\sigma$ is the index for discrete degrees of freedom such as spin and valleys. Using the fact $\hat\rho(\mathbf q)=\sum_{\mathbf k,\sigma}|\mathbf k-\mathbf q,\sigma\rangle\langle\mathbf k,\sigma|$, we can obtain the matrix elements
\begin{equation}
    \langle\psi_{\mathbf k-\mathbf q},m|\hat\rho(\mathbf q)|\psi_{\mathbf k},m\rangle=\sum_{\mathbf G,\sigma}\mu_\sigma
^\ast\left(m,\mathbf k-\mathbf q,\mathbf G\right)\mu_\sigma\left(m,\mathbf k,\mathbf G\right).
\end{equation}
Here we extend the definition of the Moir\'e Bloch state $|\psi_{\mathbf k-\mathbf q},m\rangle$ from the Moir\'e BZ to $\mathbb{R}^2$, so that $\mathbf G$ can be treated as a dummy index and the momentum in the interaction can conveniently take all values in $\mathbb{R}^2$. With these notations, we can write the form factor for Moir\'e systems as
\begin{equation}
    \rho(\mathbf k,\mathbf q)=\sum_{\mathbf G,\sigma}\mu^\ast_\sigma\left(m,\mathbf k-\mathbf q,\mathbf G\right)\mu_\sigma\left(m,\mathbf k,\mathbf G\right).
\end{equation}
Now the vector $\mu$ has its components labelled by $\mathbf G$ and $\sigma$. As in the main text, we will not write out the index $m$ in later computation. We will also neglect the the index $\sigma$ when we are emphasizing the different properties brought by Moir\'e patterns. After these simplifications of notations, the Moir\'e Bloch states defined on $\mathbb{R}^2$ satisfy the following equivalence relations:
\begin{equation}
    |\psi_{\mathbf k+\mathbf G}\rangle=e^{i\phi_{\mathbf k,\mathbf G}}|\psi_{\mathbf k}\rangle,\quad \mu(\mathbf k+\mathbf G,\mathbf G')=e^{i\phi_{\mathbf k,\mathbf G}}\mu(\mathbf k,\mathbf G+\mathbf G'),\quad c^\dagger_{\mathbf k+\mathbf G}=e^{i\phi_{\mathbf k,\mathbf G}}c^\dagger_{\mathbf k}.\label{eq_eqrlm}
\end{equation}
A central difference here from usual lattice systems is that, in addition to the Berry phase, the components of $\mu(\mathbf k+\mathbf G)$ are shifted by $\mathbf G$ compared to $\mu(\mathbf k)$. It is more convenient to view the Bloch vector $\mu(\mathbf k,\mathbf G)$ as a map from $\mathbb{R}^2$ to $\mathbb{C}P^{N-1}$.

\section{Hole Hamiltonian}

We perform a particle-hole transformation $c_{\mathbf k}\to d^\dagger_{-\mathbf k}$ to the projected Hamiltonian \eqref{eq_smpjH}. This transformation is most appropriate when the flat band is completely filled. The interaction needs to be normal-ordered so that it acts on the two-body space and annihilates the reference state, the completely filled state. In doing so, we obtain two kinetic terms for holes, by commuting those $d^\dagger,d$ terms from the same density operators or those from different density operators in Eq.~\eqref{eq_smpjH}. The former is like a mean-field uniform background contribution, denoted as $\overline\varepsilon_{\mathbf k}$, while the latter is an exchange contribution denoted as $\widetilde\varepsilon_{\mathbf k}$. For a non-Moir\'e lattice system, the background contribution is given by
\begin{equation}
   \sum_{\mathbf Q,\mathbf k'}-V(\mathbf Q)\rho(\mathbf k',-\mathbf Q)\rho(\mathbf k,\mathbf Q)\{c^\dagger_{\mathbf k'+\mathbf Q},c_{\mathbf k'}\}d_{\mathbf k}c^\dagger_{\mathbf k-\mathbf Q}=\sum_{\mathbf Q,\mathbf k'}-V(\mathbf Q)c_{\mathbf k}c^\dagger_{\mathbf k}=\overline\varepsilon d^\dagger_{-\mathbf k}d_{-\mathbf k},\label{eq_sndc}
\end{equation}
where in the second equality we use the property $\rho(\mathbf k,\mathbf Q)c^\dagger_{\mathbf k-\mathbf Q}=\rho(\mathbf k,0)c^\dagger_{\mathbf k}$ and $\rho(\mathbf k,0)=1$. This term is non-dispersing and does not change the spectrum of the Hamiltonian. We can simply drop the $\mathbf k$-dependence and write $\overline\varepsilon=-\sum_{\mathbf Q}V(\mathbf Q)$. The second exchange term is expressed as
\begin{equation}
    \sum_{\mathbf q}V(\mathbf q)\rho(\mathbf k,\mathbf q)\rho(\mathbf k-\mathbf q,-\mathbf q)c_{\mathbf k}c^\dagger_{\mathbf k}=\sum_{\mathbf q}V(\mathbf q)\left|\rho(\mathbf k,\mathbf q)\right|^2c_{\mathbf k}c^\dagger_{\mathbf k}=\widetilde\varepsilon_{-\mathbf k}d^\dagger_{-\mathbf k}d_{-\mathbf k},
    \label{eq_focklike}
\end{equation}
this term has an explicit dispersion and gives the holes a non-trivial kinetic energy.

The above formulae constructed for usual lattice systems need to be modified for Moir\'e superlattices. When extending the definition of the Bloch states to $\mathbb{R}^2$, the Moir\'e Bloch vectors $\mu$ need a further cyclic transformation which affects the inner product $\sum_{\mathbf G} \mu^\ast(\mathbf k',\mathbf G)\mu(\mathbf k,\mathbf G)$. As a result, the absolute value of the form factor is no longer periodic $|\rho(\mathbf k,\mathbf q+\mathbf G)|\ne |\rho(\mathbf k,\mathbf q)|$. We cannot move all reciprocal lattice vectors to the origin as done in \eqref{eq_sndc}. With the help of Eq.~\eqref{eq_eqrlm}, the mean-field background may also contribute a dispersion:
\begin{align}
   \sum_{\mathbf G,\mathbf k'}-\frac{V(\mathbf G)}{2}\rho(\mathbf k',-\mathbf G)\rho(\mathbf k,\mathbf G)\{c^\dagger_{\mathbf k'+\mathbf G},c_{\mathbf k'}\}c_{\mathbf k}c^\dagger_{\mathbf k-\mathbf G}+\mathrm{H.C.}=&\sum_{\mathbf G,\mathbf k'}-\frac{V(\mathbf G)}{2}\rho(\mathbf k',-\mathbf G)\rho(\mathbf k,\mathbf G)e^{i(\phi_{\mathbf k',\mathbf G}+\phi_{\mathbf k,-\mathbf G})}c_{\mathbf k}c^\dagger_{\mathbf k}+\mathrm{H.C.}\nonumber\\
   & =  \overline\varepsilon_{-\mathbf k} d^\dagger_{-\mathbf k}d_{-\mathbf k},
   \label{eqn:fullhartreelike}
\end{align}
where $\mathrm{H.C.}$ is for the Hermitian conjugation. The exchange contribution needs a similar treatment. 
\begin{align}
    \sum_{\mathbf q}\frac{V(\mathbf q)}{2}\rho(\mathbf k,\mathbf q)\rho(\mathbf k-\mathbf q+\mathbf G,-\mathbf q)\{c_{\mathbf k-\mathbf q+\mathbf G},c^\dagger_{\mathbf k-\mathbf q}\}c_{\mathbf k}c^\dagger_{\mathbf k+\mathbf G}+\mathrm{H.C.},
\end{align}
where $\mathbf G$ is the unique Moir\'e reciprocal lattice vector bringing $\mathbf k-\mathbf q+\mathbf G$ back to the first Moir\'e BZ. Using Eq.~\eqref{eq_eqrlm}, we obtain
\begin{equation}
    \sum_{\mathbf q}\frac{V(\mathbf q)}{2}\rho(\mathbf k,\mathbf q)\rho(\mathbf k-\mathbf q+\mathbf G,-\mathbf q)e^{i(\phi_{\mathbf k,\mathbf G}-\phi_{\mathbf k-\mathbf q,\mathbf G})}c_{\mathbf k}c^\dagger_{\mathbf k}+\mathrm{H.C.} =   \widetilde \varepsilon_{-\mathbf k}d^\dagger_{-\mathbf k}d_{-\mathbf k}.
    \label{eqn:fullfocklike}
\end{equation}

\section{Self-consistent Hartree-Fock theory for Emergent Fermi liquids}

Due to the non-uniform band geometry, interacting Fermi liquids become a natural candidate for topological flat bands. Its occupation pattern can be worked out in the standard self-consistent Hartree-Fock approximation. As we shall see, the kinetic energy for holes can be regarded as the extreme situation of a self-consistent Fermi liquid when the band is almost filled $\nu\simeq 1$. When the filling is smaller than $1$, the electrons will first fill the areas where the trace of the metric is small, so that their exchange interaction lowers their energy.

For simplicity, we take the non-Moir\'e model for the calculation in this section. The Hartree-Fock Hamiltonian for Fermi liquids is given by
\begin{align}
     H_{HF}=&\sum_{\mathbf q,\mathbf k,\mathbf k'}V(\mathbf q)\rho(\mathbf k,\mathbf q)\rho(\mathbf k',-\mathbf q)\left(\langle c^\dagger_{\mathbf k-\mathbf q}c_{\mathbf k}\rangle c^\dagger_{\mathbf k'+\mathbf q}c_{\mathbf k'}-\langle c^\dagger_{\mathbf k-\mathbf q}c_{\mathbf k'}\rangle c^\dagger_{\mathbf k'+\mathbf q}c_{\mathbf k}\right)-\textrm{Const.}\nonumber\\
    =&\sum_{\mathbf q,\mathbf k,\mathbf k'}\left[V(\mathbf q)\rho(\mathbf k,\mathbf q)\rho(\mathbf k',-\mathbf q)-V(\mathbf q+\mathbf k'-\mathbf k)\rho\left(\mathbf k',\mathbf q+\mathbf k'-\mathbf k\right)\rho\left(\mathbf k,-\mathbf q+\mathbf k-\mathbf k'\right)\right]\langle c^\dagger_{\mathbf k-\mathbf q}c_{\mathbf k}\rangle c^\dagger_{\mathbf k'+\mathbf q}c_{\mathbf k'}-\textrm{Const. }\nonumber\\
    \equiv&\sum_{\mathbf k',\mathbf Q} H(\mathbf k',\mathbf Q)c^\dagger_{\mathbf k'+\mathbf Q}c_{\mathbf k'}-\textrm{Const. },
\end{align}
where from the second to the third line we assume only the expectation value $\langle c^\dagger_{\mathbf k-\mathbf Q}c_{\mathbf k}\rangle$ with $\mathbf Q$ being the reciprocal lattice vectors is non-zero. For the Fermi-liquid, we have $\langle c^\dagger_{\mathbf k}c_{\mathbf k}\rangle=1$ for the $\mathbf k$ inside the Fermi surface and $\langle c^\dagger_{\mathbf k}c_{\mathbf k}\rangle=0$ for those  $\mathbf k$ outside the Fermi surface. As before, the summations over $\mathbf k,\mathbf k'$ are limited to the Brillouin zone. Using the identities in Eq.~\eqref{eq_eqBZtrQ}, we can further simplify to 
\begin{equation}
     H_{HF}=\sum_{\mathbf k}E_{\mathbf k}c^\dagger_{\mathbf k}c_{\mathbf k}-\textrm{Const. },\quad E_{\mathbf k}=\sum_{\mathbf k',\mathbf Q}\left[V(\mathbf Q)-V(\mathbf Q+\mathbf k'-\mathbf k)\rho(\mathbf k,\mathbf k-\mathbf k')\rho(\mathbf k',\mathbf k'-\mathbf k)\right]\langle c^\dagger_{\mathbf k'}c_{\mathbf k'}\rangle,
     \label{eq_fullHFenergy}
\end{equation}
As before, the first term in $E_{\mathbf k}$ is not dispersing and can be thrown away. For the second term in $E_{\mathbf k}$, we can use the summation over $\mathbf Q$ to extend the definition of $\mathbf k'$ to $\mathbb{R}^2$. After doing a change of variable $\mathbf q=\mathbf k-\mathbf k'$, we obtain
\begin{equation}
    E_{\mathbf k}=\sum_{\mathbf Q}\nu V(\mathbf Q)-\sum_{\mathbf q}V(\mathbf q)|\rho(\mathbf k,\mathbf q)|^2\langle c^\dagger_{\mathbf k-\mathbf q}c_{\mathbf k-\mathbf q}\rangle=\sum_{\mathbf Q}\nu V(\mathbf Q)-\sum_{\mathbf q}V(\mathbf q)|\rho(\mathbf k,\mathbf q)|^2f(E_{\mathbf k-\mathbf q}-u),\label{eq_slfcFL}
\end{equation}
where in the second step we use the Fermi-Dirac distribution function $f$ to replace the expectation value. The chemical potential $u$ corresponds to the  energy at the Fermi surface. The filling of the electrons is denoted by $\nu$. Notice that the energy $E_{\mathbf k}$ appears on both sides, this equation needs to be solved self-consistently. This can be done via iterations. One starts with some trial Fermi-liquid ansatz $\langle c^\dagger_{\mathbf k}c_{\mathbf k}\rangle$ and finds a trial dispersion $E^{(0)}_{\mathbf k}$. Then choose a chemical potential $u^{(0)}$ so that particle number is at the desired value $\sum_{\mathbf k}f(E^{(0)}_{\mathbf k}-u^{(0)})=\nu$. Inserting $u^{(0)}$ and $E^{(0)}_{\mathbf k}$ into the right side of \eqref{eq_slfcFL}, we obtain the dispersion of the first iteration $E^{(1)}_{\mathbf k}$ and $u^{(1)}$ again needs to be chosen so that $\sum_{\mathbf k}f(E^{(1)}_{\mathbf k}-u^{(1)})=\nu$. The final result is obtained by repeating this process until $E^{(m)}_{\mathbf k}$ becomes stable at some $m$.

Eq.~\eqref{eq_slfcFL} tells us how electrons occupy the Brillouin zone when we start to fill the band from $\nu=0$. Using the approximation in the main text, the self-consistent energy of electrons can be expressed by (we neglect the constant part here)
\begin{equation}
     E_{\mathbf k}\simeq-\sum_{\mathbf q}V(\mathbf q)e^{-\sum_{ab} q_a q_bg_{ab}(\mathbf k)}\langle c^\dagger_{\mathbf k-\mathbf q}c_{\mathbf k-\mathbf q}\rangle.
\end{equation}
Assuming that $g_{ab}(\mathbf k)$ is a smooth but non-constant function in $\mathbf k$, this Fock energy is maximally lowered when $\textrm{tr }g(\mathbf k)$ takes its smallest value, say at $\mathbf k_\ast$, in the Brillouin zone. As a result, the electrons prefer to accumulate around $\mathbf k_\ast$ and benefit from their lowest Fock energy. As we increase the filling, the electrons always seek to find themselves in the area with strongest Fock interaction. i.e. smallest  $\textrm{tr }g(\mathbf k)$. In this sense, the Fubini-Study metric guides the electrons to form a Fermi-liquid state.

To connect this result to the hole energy, we look at the special situation when the band is completely filled. In this situation we simply need to choose a sufficiently small $u$ so that the Fermi-Dirac distribution function $f(E_{\mathbf k}-u)=1$ for all $\mathbf k$.  Compared to the expression of the hole energy \eqref{eq_focklike}, one immediately observes that the hole kinetic energy is equal to the Hartree-Fock energy of the electrons at $\nu=1$
\begin{equation}
    -\varepsilon_{-\mathbf k}=E_{\mathbf k}\big\vert_{\nu=1}
\end{equation}
This is to say, the hole kinetic energy can be understood as an electron receiving the Fock interaction from all other electrons in the Brillouin zone. When $\nu<1$, the electron only receives the Fock energy from the occupied areas of the Brillouin zone. The self-consistent electron energy is slightly higher than the hole energy $E_{\mathbf k}\ge-\varepsilon_{-\mathbf k}$ (we neglect the constant Hartree energy again). However, as we will see, the Moir\'e form factor is decaying quickly in momentum space. The main contribution to the Fock energy $E_{\mathbf k}$ are coming from those $\mathbf k'$ close to it, $\mathbf k'\sim \mathbf k$. So for the electrons living away from the Fermi surface, the hole dispersion $-\varepsilon_{-\mathbf k}$ gives an accurate estimate of $E_{\mathbf k}$. Only a small deviations happens when the electron is exactly at the Fermi surface. We give an example for this in Fig. \ref{fig_HFvsHole} for $\nu = 1/2$ in the $C = 3$ TLG-hBN band (see Fig.2.(d) in the main text). The self-constant energy only differs from the hole energy by a very small correction. In sum, the bulk of the Fermi-liquid in the Brillouin zone can be well located by simply looking at the hole dispersion.

\begin{figure}
    \centering
    \includegraphics[width=0.85\linewidth]{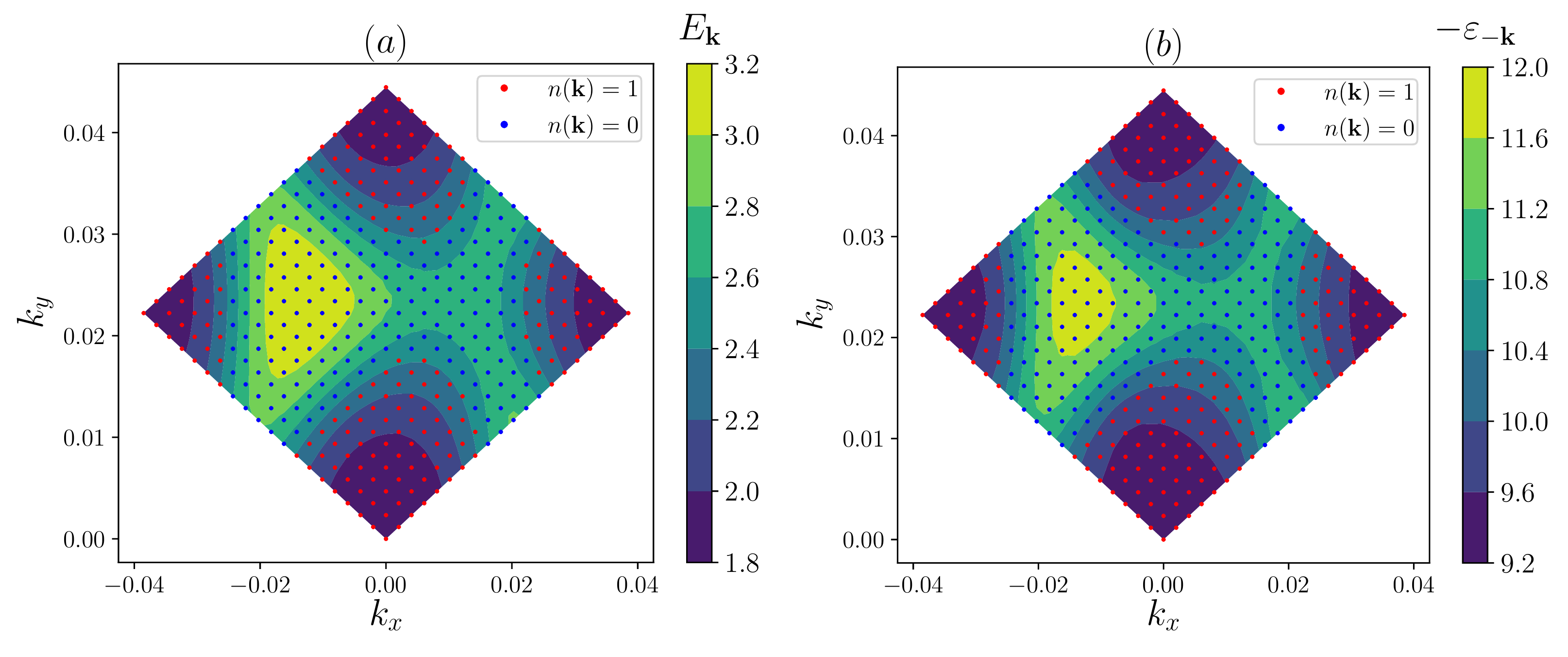}
    \caption{Self-consistent  Fermi liquid Hartree-Fock energy $E_{\mathbf{k}}$ obtained from equation \eqref{eq_fullHFenergy} (shown in (a)) vs the hole energy $\varepsilon_{\mathbf{-k}}$ (shown in (b)) for the $C=3$ TLG-hBN band  at $\nu = 1/2$. Plotted on top in both figures is the occupation $n(\mathbf{k})$ obtained from the self-consistent calculation. i.e, filling the states with the lowest $E_{\mathbf{k}}$. Both $E_{\mathbf{k}}$ and $\varepsilon_{\mathbf{-k}}$ look very similar away from the Fermi surface while there is a slight deviation in the occupation around the Fermi surface. } 
    \label{fig_HFvsHole}
\end{figure}

\section{Effect of residual terms on the single-hole dispersion}

Equations \eqref{eqn:fullhartreelike} and \eqref{eqn:fullfocklike} define the full single-hole energy obtained from the particle-hole transformation in the case of Moir\'e systems where the the form factors are no longer periodic.  For decaying potentials $V(\mathbf{q}) \sim 1/|q|$ and decaying form factors (c.f Fig. \ref{fig_formfactordecay}), we can drop out the sum over $\mathbf{G}$ in equations \eqref{eqn:fullhartreelike} and in this case, we end up with a non dispersive $\overline{\varepsilon}_{-\mathbf{k}}$ that we can simply neglect. As for the exchange interaction term \eqref{eqn:fullfocklike}, it assumes a similar form to equation \eqref{eq_focklike} when the momentum transfer is such that $\mathbf{k}-\mathbf{q} \in $ the Moir\'e Brilloin zone (MBZ). This further allows the expansion in terms of the Fubini-Study metric in the limit of small momentum transfer $\mathbf{q}$ as shown in equation (5) in the main text therefore equations \eqref{eqn:fullfocklike} and \eqref{eq_focklike} are identical in the limit of small $\mathbf{q}$. When the momentum transfer is such that $\mathbf{k}-\mathbf{q} \not\in
 \rm MBZ$, equation \eqref{eqn:fullfocklike} and \eqref{eq_focklike} are generally different. However because of the suppression by the decaying interaction potential and form factors, the contribution from momenta transfer $\mathbf{k}-\mathbf{q} \not\in
 \rm MBZ$ is expected to be small. 
 
To illustrate this, let's define $\widetilde \varepsilon_{-\mathbf{k}}^{\rm approx}$ as \begin{equation} 
 \widetilde \varepsilon_{-\mathbf{k}}^{\rm approx} = 
     \sum_{\mathbf q : \> \mathbf{k}-\mathbf{q} \in \rm MBZ }V(\mathbf q)\left|\rho(\mathbf k,\mathbf q)\right|^2
     \label{eq_approxexchange}
 \end{equation}
 Where we truncate the sum in \eqref{eqn:fullfocklike} such that $\mathbf{k}-\mathbf{q} \in \rm MBZ$. We calculate $W_{\overline{\varepsilon}}$, $W_{\widetilde{\varepsilon}}$,  $W_{\widetilde{\varepsilon}}^{\rm approx}$ and  $W$ that we define as the bandwidth of the mean-field contribution \eqref{eqn:fullhartreelike}, the full exchange contribution \eqref{eqn:fullfocklike}, the truncated exchange contribution \eqref{eq_approxexchange}  and the full hole dispersion (\eqref{eqn:fullhartreelike} + \eqref{eqn:fullfocklike}) respectively. The results are summarized in table \ref{Tab:bandwidth} for TLG-hBN.  It's clear that the mean field contribution to the bandwidth $W_{\overline{\varepsilon}}$ is negligible and it's the exchange interaction \eqref{eqn:fullfocklike} that dominates the hole dispersion. A major contribution to the exchange interactions comes from momentum transfer $\mathbf{k}-\mathbf{q} \in \rm MBZ$. 
 
 In addition to that, we show that the occupation $n(\mathbf{k})$ correlates well with the truncated exchange interaction \eqref{eq_approxexchange} obtained when $\mathbf{k}-\mathbf{q} \in \rm MBZ$ in a similar manner as what we show in Fig 2. in the main text as shown in Fig. \ref{fig_approxfock}. This further confirms the expectation that the effect of the single-hole dispersion could be still captured by equation \eqref{eq_focklike}.
 \begin{figure}[h!]
    \centering
    \includegraphics[width=0.85\linewidth]{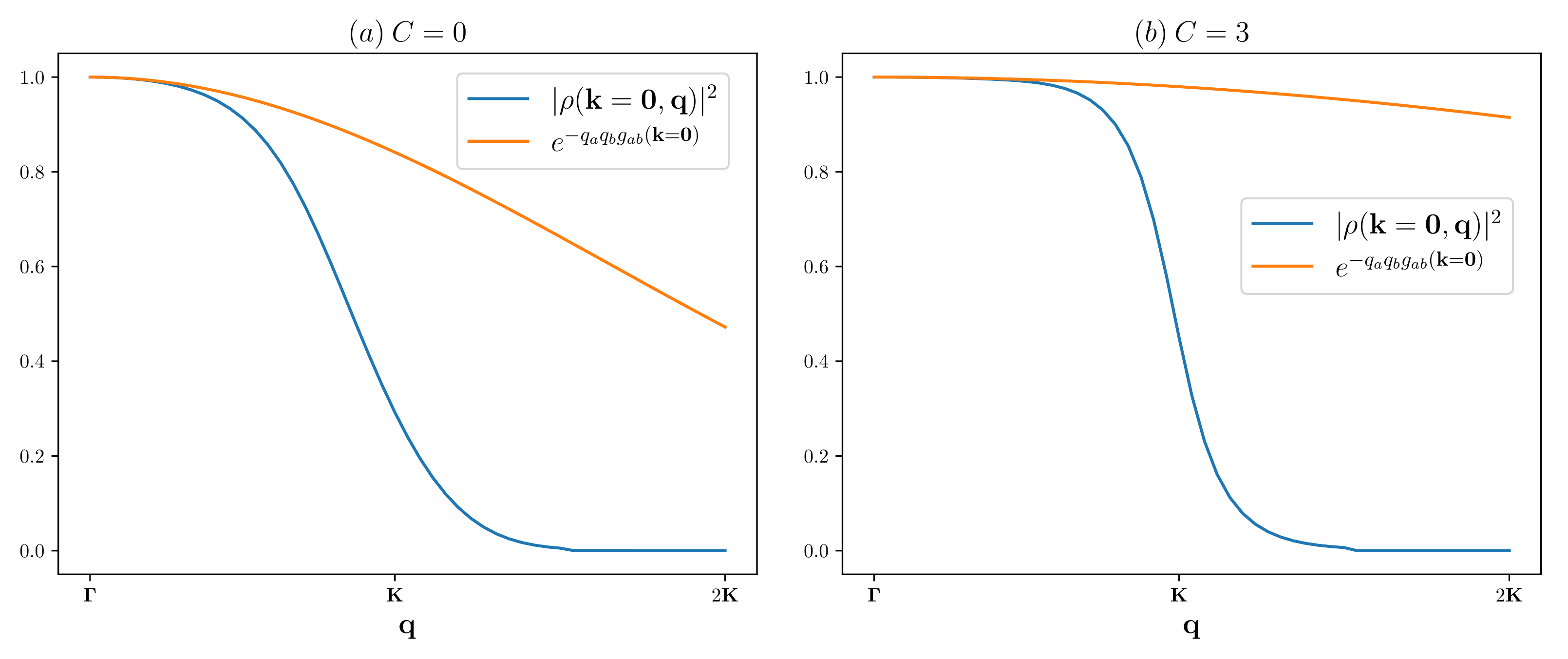}
    \caption{Form factor decay in TLG-hBN for (a) $C = 0$ band and (b) $C = 3$ band: The momentum $\mathbf{k}$ is fixed to be at the $\Gamma$ point while the $\mathbf{q}$ values are swept along a line that extends from the $\Gamma$ point to $2 \mathbf{K}$ point that lie outside the Moir\'e Brillouin zone. The form factor decay is compared against the exponential approximation of the form factors in equation (4) of the main text.} 
    \label{fig_formfactordecay}
\end{figure}
{\setlength{\tabcolsep}{10pt}
\renewcommand{\arraystretch}{1.5}
\begin{center}
\begin{table}
\begin{tabular}{| p{2.5cm}| p{1cm}| p{1cm}| p{2cm}|}
 
 \hline
  & $W_{\overline{\varepsilon}}/{W}$  &  $W_{\widetilde{\varepsilon}}/{W}$ &  $W_{\widetilde{\varepsilon}}^{\rm approx}/{W}$  \\ 
  \hline
{TLG-hBN $C = 0$} & 0.0941 & 0.9136 & 0.6171 \\ 
 \hline
 TLG-hBN $C = 3$ & 0.0454 & 0.9845 & 0.6626 \\ 
 \hline
\end{tabular}
\caption{Comparing the bandwidth of the mean-field contribution \eqref{eqn:fullhartreelike}, the exchange contribution \eqref{eqn:fullfocklike} and the  exchange contribution  \eqref{eqn:fullfocklike} when  $\mathbf{k}-\mathbf{q} \in \rm MBZ$ to the bandwidth $W$ of the full single-hole dispersion.}
\label{Tab:bandwidth}
\end{table}
\end{center}}
\begin{figure}[h!]
    \centering
    \includegraphics[width=0.85\linewidth]{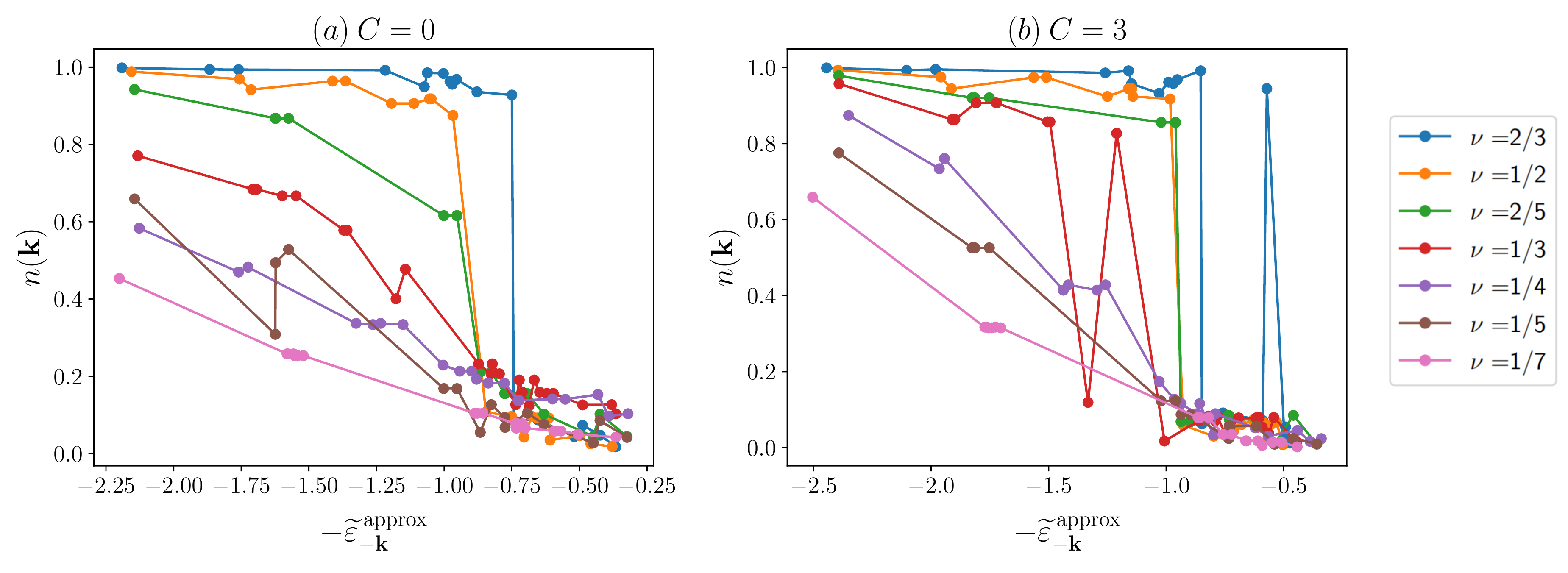}
    \caption{Ground state electron occupation $n(\mathbf{k}) = \langle c^\dagger_{\mathbf{k}} c_{\mathbf{k}} \rangle$ vs the truncated exchange interaction $ \widetilde \varepsilon_{-\mathbf{k}}^{\rm approx}$ defined in equation \eqref{eq_approxexchange} for TLG-hBN (a) $C=0$ band and (b) $C=3$ band.} 
    \label{fig_approxfock}
\end{figure}

\section{Some approximated results for hole dispersion}

In this section, we obtain some analytic results for the hole dispersion and show that the hole kinetic energy is mainly controlled by the trace of the metric. For simplicity, we assume the interaction is rotationally invariant. With the quadratic approximation to the form factor, the hole kinetic energy depends on two invariant quantities of the Fubini-Study metric, its trace $\textrm{tr }g$ and its determinant $\textrm{det }g$. These two quantities satisfy the inequality $(\textrm{tr }g)^2\ge 4\textrm{det }g$ and the equality holds when the metric is isotropic.

We look at the situation when the interaction potential has a cutoff in momentum space. For the Zhang-Das Sarma potential with large $l$, we can use the quadratic approximation. The result is
\begin{equation}
      \widetilde\varepsilon_{-\mathbf k}\simeq \sum_{\mathbf q} \frac{(2\pi e^2)e^{-ql}}{q}\left[1-\sum_{ab} q_a q_bg_{ab}(\mathbf k)\right] =\frac{e^2}{ l}\left(1-\frac{\textrm{tr }g(\mathbf k)}{l^2}\right).
\end{equation}
When the cutoff $l$ is not very small, we need to resort to the exponential approximation for the form factor. The result is very complicated when $g_{ab}(\mathbf k)$ is anisotropic. We only give the analytic result for an isotropic Fubini-Study metric:
\begin{equation}
    \widetilde\varepsilon_{-\mathbf k}\simeq\sum_{\mathbf q} \frac{(2\pi e^2) e^{-ql}}{q}e^{-\sum_{ab} q_a q_bg_{ab}(\mathbf k)} =e^2\sqrt{\frac{\pi}{2\textrm{tr }g(\mathbf k)}}e^{\frac{l^2}{2\textrm{tr }g(\mathbf k)}}\textrm{erfc}\left(\frac{l}{\sqrt{2\textrm{tr }g(\mathbf k)}}\right),\quad, \textrm{for $(\textrm{tr }g)^2\simeq 4\textrm{det }g$ },
\end{equation}
where $\textrm{erfc}(x)$ is the  complementary error function.

\begin{figure}
    \centering
    \includegraphics[width=0.5\linewidth]{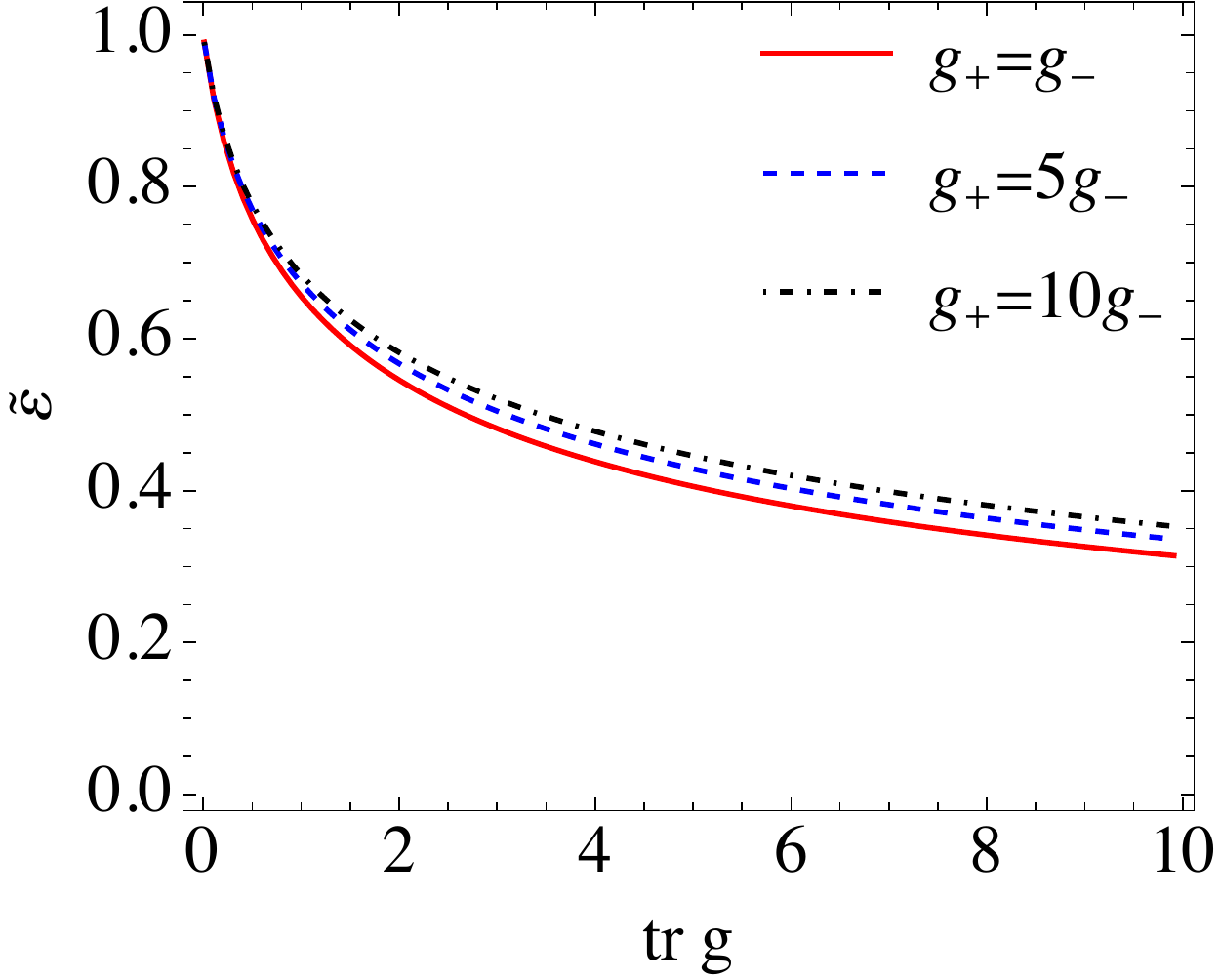}
    \caption{The dependence of the approximated hole exchange energy on the trace of the metric and the ratio between its two eigenvalues $g_+/g_-$. The Zhang-Das Sarma potential \cite{PhysRevB.33.2903} with a cutoff length $l$ is employed. The unit of the metric is taken as $l^2$ and the unit for the energy is $ e^2/(l\epsilon)$ with $\epsilon$ the dielectric constant.}
    \label{fig_ani_mtc}
\end{figure}

Through numerical calculations for different anisotropic metrics, we find that the trace of the metric plays the most significant role in estimating the hole energy, while anistropy brings small corrections. We denote the two eigenvalues of the metric as $g_+$ and $g_-$ and use $l$ as the unit of length. We compare the kinetic energy at the isotropic case $g_+=g_-$ with the highly anisotropic case $g_+=10g_-$. The result is summarized in Fig.~\ref{fig_ani_mtc}. The hole energy can be treated as a nearly monotonic function of $\textrm{tr }g$.

\section{Exact Diagonalization Details}

Through out the paper, we use exact diagonalization techniques to compute quantities such that the many-body spectrum, the electron occupation $n(\mathbf{k})$ and the static structure factor $S(\mathbf{q})$. Our starting point is the interacting Hamiltonian projected onto a flat band. It reads \begin{align}
    H=&\frac{1}{2}\sum_{\mathbf q ,\mathbf k,\mathbf k'}V(\mathbf q) \rho(\mathbf{k},\mathbf{q}) \rho(\mathbf{k}',-\mathbf{q}) c^\dagger_{\mathbf {k-q}}c^\dagger_{\mathbf {k'+q}} c_{\mathbf k'} c_{\mathbf k}.
\end{align} 
We diagonalize the above Hamiltonian on a finite lattice. Since we address the possibility of stabilizing charge density waves that are sensitive to the geometry of the lattice, the choice of the finite lattice becomes important. It's crucial to choose finite lattices that are commensurate with the Moir\'e triangular lattice in inspecting possible CDWs. We follow a technique that was used in \cite{PhysRevLett.111.126802} and \cite{PhysRevB.103.125406} that we summarize here again. The finite lattice is spanned  by the two vectors $\mathbf{T}_1$ and $\mathbf{T}_2$  \begin{align}
    \mathbf{T}_1 = a \>  \mathbf{a}_1 + b \> \mathbf{a}_2  \nonumber \\
    \mathbf{T}_2 = c \> \mathbf{a}_2 + d \> \mathbf{a}_2 
    \label{eq_finitelattice}
\end{align} With $a$,$b$, $c$ and $d$ being integers and $\mathbf{a}_1$ and $\mathbf{a}_2$ are the two Moir\'e lattice vectors. Particular choices of $a$,$b$, $c$ and $d$ result in finite lattices with the required symmetry. By imposing periodic boundary conditions, it's possible to extract the momentum space discretization basis $\mathbf{g}_1 $ and $\mathbf{g}_2$ such that a momentum point is given by $\mathbf{k} = m \mathbf{g}_1 + n \mathbf{g}_2$ for integer $m$ and $n$. The electron filling factor is then given by $\nu = N_e/N_k$ where $N_e$ and $N_k$ are the number of band electrons and total number of momentum points respectively. 
\begin{figure}
    \centering
    \includegraphics[width=0.85\linewidth]{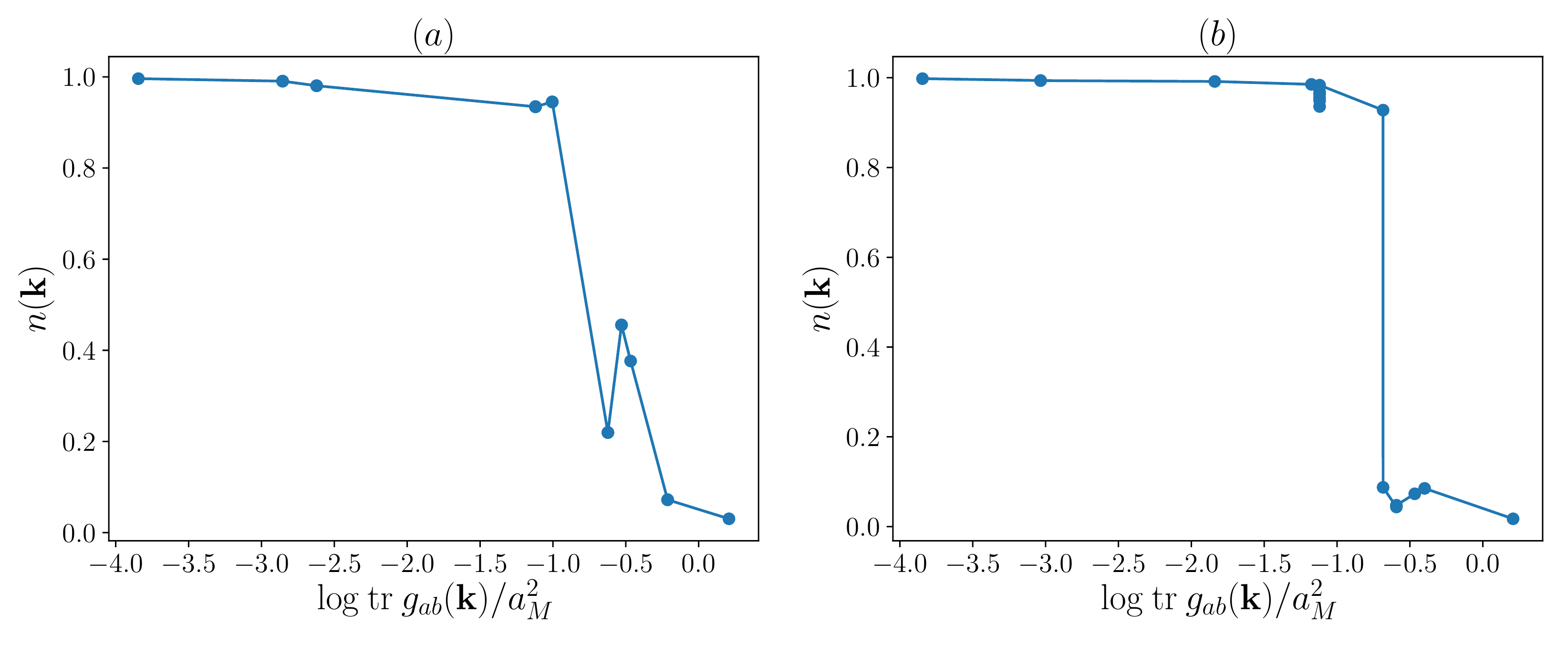}
    \caption{TLG-hBN $C= 0$ at $\nu =2/3$: (a) Electron occupation $n(\mathbf{k})$  when averaged over the lowest 15 states shown in Fig.~3(c) (of the main text) vs $\textrm{log tr }g_{ab}(\mathbf{k})/a_M^2$. (b) Electron occupation $n(\mathbf{k})$ in the lowest energy ground state calculated for a finite system with $a = 3, b = 0, c = 0, d= 6$ (see equation \eqref{eq_finitelattice}) vs $\textrm{log tr }g_{ab}(\mathbf{k})/a_M^2$.}
    \label{fig_twothirdanalysis}
\end{figure}

\section{Analysis of the \texorpdfstring{$\nu = 2/3$}{nu=2/3} in TLG-hBN C = 0}
In the main text, we argued that the strong metric fluctations in TLG-hBN results in a strong particle-hole asymmetry that destroys a possible CDW at the particle-hole dual filling $\nu = 2/3$. As shown in Fig. 2(c). in the main text, the occupation $n(\mathbf{k})$ when calculated in the lowest energy ground state exhibits oscillations around the Fermi surface for filling $\nu = 2/3$. Here we argue that these oscillations are most likely resulting from finite-size effects. If the many-body spectrum exhibits a number of quasi-degenerate ground states as in the case shown in Fig. 3(c). in the main text, these oscillations could be smoothed out by considering not only the lowest ground state but a number of low-lying states. Since we are considering an interacting problem, interactions could mix the non-interacting Fermi liquid ground states (obtained by filling the highest hole energies) among themselves and also with other low-lying states. For numerics on finite lattices, this is enhanced if the underlying lattice contains lots of points around the Fermi energy.  To demonstrate this, we average the occupation $n(\mathbf{k})$ over a number of low lying many-body states and find that oscillations are smoothed out to a great extent as shown in Fig. \ref{fig_twothirdanalysis}(a). In addition, we consider a smaller system that has fewer points around the Fermi energy and find that the occupation in the lowest energy state to exhibit a sharp jump as shown in Fig. \ref{fig_twothirdanalysis}(b). This confirms that this state is very likely to be a Fermi liquid state.
\section{FCI vs CDW in twisted bilayer graphene}
\begin{figure}[t!]
    \centering
    \includegraphics[width=1\linewidth]{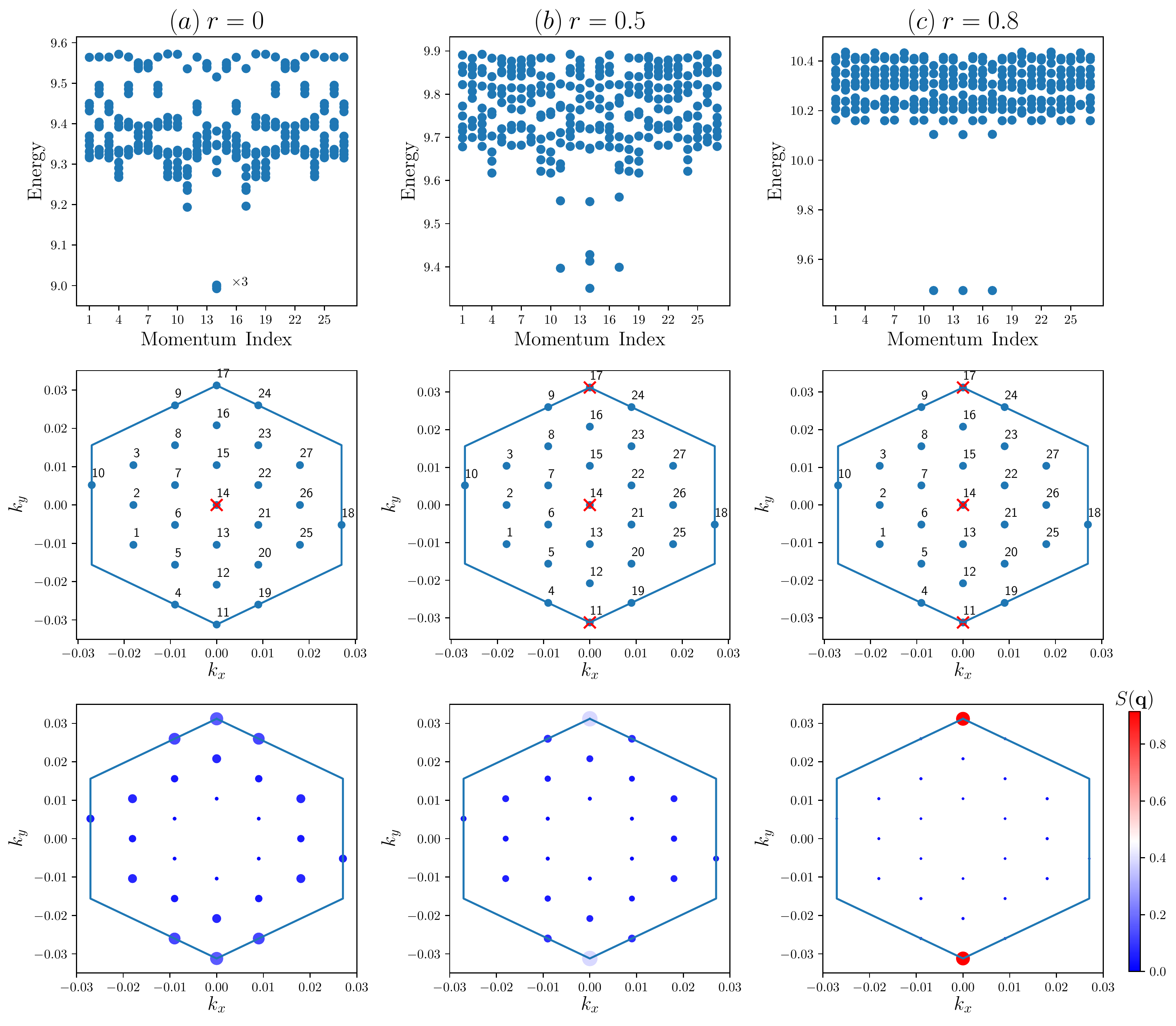}
    \caption{Exact diagonalization results at $\nu = 1/3$ of the spin-valley polarized valence band in TBG-hBN for different values of $r = w_0/w_1$. Upper panel: many-body spectrum vs total momentum. Middle panel: location of the ground state total momenta marked by red $\times$ in the finite lattice. Lower panel: static structure constant $S(\mathbf{q}) = \langle \hat{\rho}^{\rm proj}_{\mathbf{q}} \hat{\rho}^{\rm proj}_{-\mathbf{q}} \rangle $  calculated in the  many-body ground state  with $\hat{\rho}^{\rm proj}_{\mathbf{q}} = \sum_{\mathbf{k}} \mu^\dagger(\mathbf{k} - \mathbf{q}) \mu(\mathbf{k}) c^\dagger_{\mathbf{k}-\mathbf{q}} c_{\mathbf{k}}$. Simulations were performed on a finite lattice defined by $a = 3$, $b =3$, $c =3$, $d = -6$ (see equation \eqref{eq_finitelattice}) with $N_e = 9$ electrons} 
    \label{fig_EDFCIvsCDW}
\end{figure}
The compeition between FCIs and CDWs filling $\nu = 1/3$ of the flat band in twisted bilayer graphene has been thoroughly investigated in the exact diagonalization study \cite{PhysRevB.103.125406} for different values of the AB tunneling strength $w_1$ and in the DMRG study \cite{parker2021field} for different $r = w_0/w_1$. In the main text, we study this competition from the point of view of the fluctuations of the Fubini-Study metric. We supplement the results shown in the main text by the rest of the exact diagonalization results on the sample values of $r$ used, $r = 0, 0.5, 0.8$. We choose $w_1 = 110$ meV and the twist angle $\theta = 1.05^\circ$. We use the screened Yukawa potential $V(\mathbf{q}) = 2\pi/\sqrt{|\mathbf{q}|^2 + \kappa^2}$ with screening length $\kappa = 1/a_{M}$ with $a_{M}$ is the Moir\'e lattice constant. In Fig. \ref{fig_EDFCIvsCDW}, we show the exact diagonalization results. The upper panel shows the many-body spectrum while the middle panel shows where the ground-state total momenta lie in the finite sample. As $r$ is changed, the location of the ground state momenta changes. When $r = 0$, we observe 3-fold degenerate ground states at zero total momentum. This is in agreement with the derived counting rules for FCI states \cite{bernevigEmergentManybodyTranslational2012} when applied to our finite lattice. Laughlin-like states satisfy a (1,3) counting rule which imposes that the ground state admissible configurations are those that have one particle in each 3 consecutive orbitals. For the titled lattices we use, the finite lattice is not rectangular in momentum space thus in order to fold the two dimensional momenta, we use the method introduced here \cite{PhysRevLett.111.126802} that is based on defining a topological extent of the finite lattice. When $ r = 0.8$, we observe 3-fold degenerate ground state at total momenta corresponding to the zero momentum point and the two distinct $\mathbf{K}$ points in accordance to a CDW state that triples the unit cell. The transition between the two states happens roughly around $r \approx 0.5$ although more thorough numerics are required to diagnose the transition. The distinction between the FCI state and the CDW state is emphasized in the static structure factor shown in the lower panel in Fig. \ref{fig_EDFCIvsCDW}. While the structure factor for the FCI state is featureless as expected from a liquid-like state, the CDW one exhibits sharp peaks at the $\mathbf{K}$ points as expected from a state that breaks position-space translational symmetry.

\end{widetext}

\end{document}